\documentclass[10pt,journal,compsoc]{IEEEtran}

\usepackage{cite}
\usepackage{url}
\usepackage[caption=false]{subfig}
\usepackage{amsmath,amsthm,amssymb}
\usepackage{algorithmic}
\usepackage{graphicx}
\usepackage{float}
\usepackage{textcomp}
\usepackage{color}
\usepackage{mathtools}
\usepackage{mathrsfs}
\usepackage{bm}
\usepackage{balance}
\usepackage{acronym}
\usepackage{epstopdf}
\usepackage{threeparttable}
\usepackage[table]{xcolor}
\usepackage{colortbl}

\newcommand{\beq}{\begin{equation}}
\newcommand{\eeq}{\end{equation}}
\newcommand{\beqa}{\begin{eqnarray}}
\newcommand{\eeqa}{\end{eqnarray}}
\DeclareMathOperator*{\argmin}{arg\,min}

\newcommand{\lmttfont}{\fontfamily{lmtt}\selectfont}

 \newcommand{\LineSep}[1][-0.2]{%
 	\par\vspace*{\dimexpr-\baselineskip+7mm}%
 } 
  
\makeatletter  
  
\makeatletter
\def\ps@IEEEtitlepagestyle{%
	\def\@oddfoot{\mycopyrightnotice}%
	\def\@oddhead{\hbox{}\@IEEEheaderstyle\leftmark\hfil\thepage}\relax
	\def\@evenhead{\@IEEEheaderstyle\thepage\hfil\leftmark\hbox{}}\relax
	\def\@evenfoot{}%
}
\def\mycopyrightnotice{%
	\begin{minipage}{\textwidth}
		\centering \scriptsize
		Copyright~\copyright~2022 IEEE. Personal use of this material is permitted. Permission from IEEE must be obtained for all other uses, in any current or future media, including reprinting/republishing this material for advertising or promotional purposes, creating new collective works, for resale or redistribution to servers or lists, or reuse of any copyrighted component of this work in other works by sending a request to pubs-permissions@ieee.org.
	\end{minipage}
}
\makeatother
  
\begin{document}

\title{A Latency-driven Availability Assessment for \\ Multi-Tenant Service Chains}

\author{Luigi~De~Simone,~\IEEEmembership{Member,~IEEE,}
	Mario~Di~Mauro,~\IEEEmembership{Senior Member,~IEEE,}
		Roberto~Natella,
		Fabio~Postiglione
}

\IEEEtitleabstractindextext{%
\begin{abstract}
Nowadays, most telecommunication services adhere to the Service Function Chain (SFC) paradigm, where network functions are implemented via software. In particular, container virtualization is becoming a popular approach to deploy network functions and to enable resource slicing among several tenants. The resulting infrastructure is a complex system composed by a huge amount of containers implementing different SFC functionalities, along with different tenants sharing the same chain. The complexity of such a scenario lead us to evaluate two critical metrics: the steady-state availability (the probability that a system is functioning in long runs) and the latency (the time between a service request and the pertinent response). Consequently, we propose a latency-driven availability assessment for multi-tenant service chains implemented via Containerized Network Functions (CNFs). We adopt a multi-state system to model single CNFs and the queueing formalism to characterize the service latency. To efficiently compute the availability, we develop a modified version of the Multidimensional Universal Generating Function (MUGF) technique. Finally, we solve an optimization problem to minimize the SFC cost under an availability constraint. As a relevant example of SFC, we consider a containerized version of IP Multimedia Subsystem, whose parameters have been estimated through fault injection techniques and load tests.
\end{abstract}

}

\maketitle

\vspace{-35pt}
\section{Introduction}

\begin{figure}[t]
    \centering
    \captionsetup{justification=centering}
    \includegraphics[scale=0.26,angle=90]{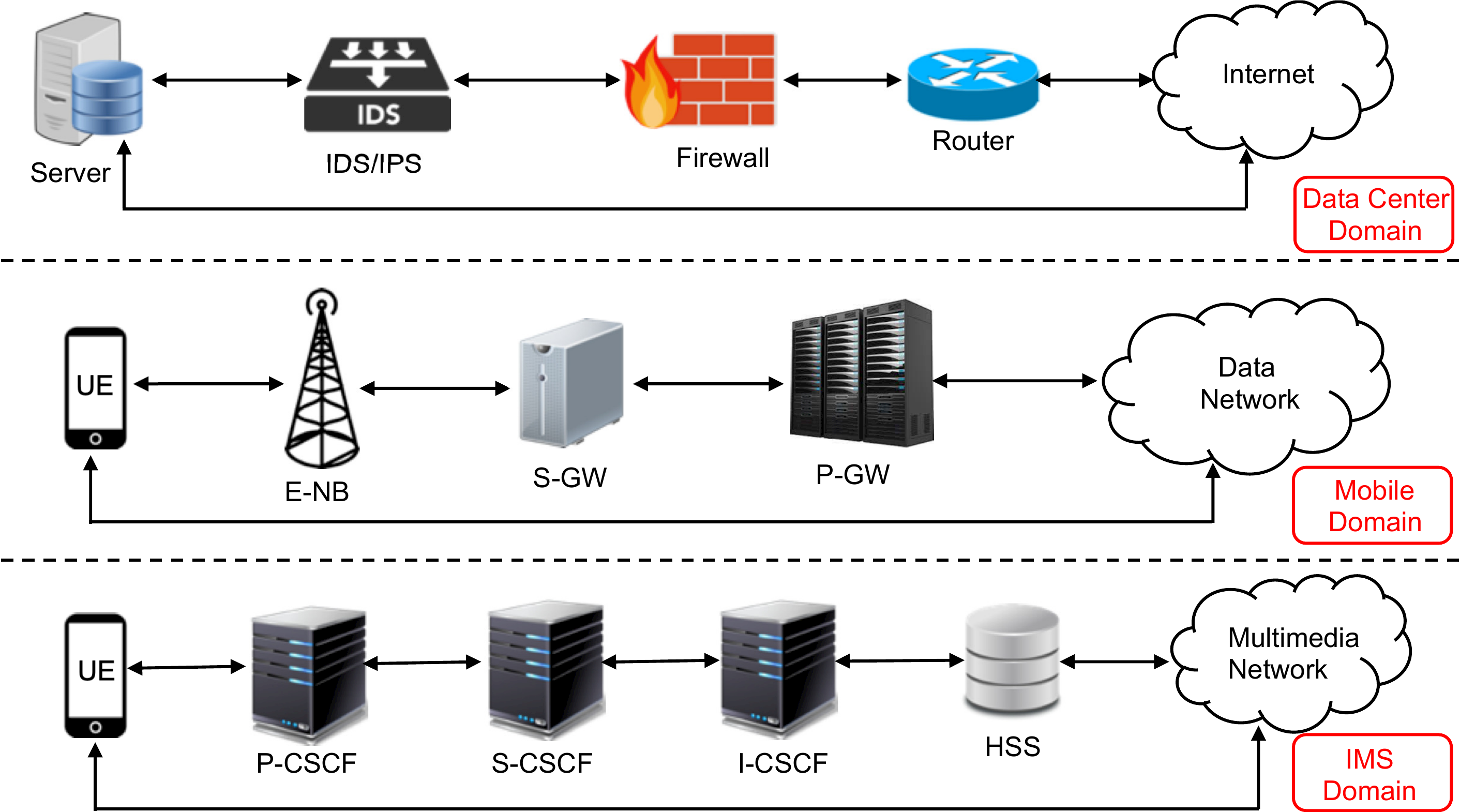}
    \caption{Examples of domains embracing the Service Chains paradigm: Data Center (upper panel), Mobile Network (middle panel), IP Multimedia Subsystem (lower panel).}
    \label{fig:sfc_domains}
\end{figure}

\IEEEPARstart{T}{oday}, service providers conceive modern network infrastructures by taking into account cloud-centric paradigms such as Network Function Virtualization (NFV), which remodels classic network nodes (routers, switches, firewalls, and others) as virtual entities called Virtualized Network Functions (VNFs). 
VNFs can be chained to realize Service Function Chains (SFCs), which represent the modern way of composing and providing new services quickly and flexibly, especially when coupled with the Software Defined Networking \cite{cerroni1,cerroni2,cerroni3}. Many applications adopt the SFC paradigm \cite{sfc-dc,sfc-mobile} with some examples shown in Fig.~\ref{fig:sfc_domains}: the Data Center domain (upper panel), where the chain is made of systems such as Intrusion Detection, Firewall, and Router in charge of processing the data flow between a Server and the Internet; the cellular domain (middle panel), where the mobile traffic is managed by a chain including: enhanced Node-B (e-NB) to handle the radio link, and Signal and Packet Gateways (S-GW, P-GW) to manage the signaling and the data content, respectively; IP Multimedia Subsystem (IMS) (lower panel) which relies on a chain of network nodes providing multimedia services. 
Across such domains, virtualization enables a flexible and efficient resource utilization, since resources can be allocated and shared among several service providers (\emph{tenants}) at a fine grain. 
Examples of multi-tenant commercial and standard-based solutions include: softwarized chains in the Evolved Packet Core domain \cite{nec}, software-based IMS shared among different providers \cite{eri2014}, and the Gateway Core Network (GWCN) for infrastructure sharing among different providers \cite{etsimultitenant}. All the aforementioned systems must satisfy quality-of-service requirements, both in terms of steady-state availability (the probability that a system is functioning in long runs, i.e., when stationary conditions are reached) and latency (the time between a service request and the pertinent response).

From a technological point of view, we are witnessing the adoption of \emph{Containarized Network Functions} (CNFs) to implement VNFs \cite{cziva2017container,cotroneo17}. 
Differently from traditional virtualization technology, containers are a lightweight solution, as they do not emulate a full computer machine, and do not run a dedicated operating system. Moreover, containers can be quickly deployed and orchestrated using dedicated management platforms (e.g., Docker \cite{docker}).
Remarkably, lightweight containers allow designers to achieve a great flexibility, in terms of a fine-grained allocation of resources among various tenants. On the other hand, the complexity of managing a huge number of container replicas could negatively affect the computational cost of many availability techniques. In addition, exploiting containerized solutions in real-time environments requires a particular attention to achieve the low latency objectives. It is the case of IMS, whose latency must be below few tens of milliseconds \cite{etsicsd,csd1,csd2,csd3}.  
Therefore, there is a need for new assessment techniques that are computationally efficient, and that can address both high availability and low latency constraints. 

Aimed at dealing with the aforementioned concerns, in this work we advance:

\vspace{3pt}
\textbf{1) A novel latency-driven modeling approach for a Containerized Network Function (CNF)}, a three-layered (Software, Docker, Infrastructure layers) structure representing the elementary block of a service chain, and which has been modeled in terms of a Multi-State System (MSS). The MSS formalism is helpful to encode the interplay among the various nested layers in a containerized node, in terms of failures and repair actions. To take into account latency, which is typically neglected in the technical literature, we enhance this representation with a \emph{delay model}, based on queues with non-exponential service times and time-varying serving facilities ruled by the failure/repair process.

\vspace{3pt}

\textbf{2) A technique for the efficient analysis of service chains composed by several interconnected CNFs}, based on the \emph{Universal Generating Function} (UGF) technique. This paper extends our previous work on a multidimensional version of UGF (MUGF) \cite{MDMTSC} that supports multi-tenant service chains, where different operators (or tenants) share the same infrastructure. In this work, we revised the MUGF technique to take into account the novel latency-based metric, in order to support the analysis over a chain of CNFs.

\vspace{3pt}

\textbf{3) A detailed case study on a containerized multi-tenant IP Multimedia Subsystem (cIMS)} platform, a service chain-like infrastructure crucial to manage multimedia content within the $5$G core network. The case study shows the feasibility of the proposed approach in the context of a relevant use case. In particular, it presents an extensive set of experiments on the \emph{Clearwater} cloud-based IMS platform \cite{clearwater}, through $i)$ load test experiments, to estimate the empirical service times, and $ii)$ fault injection experiments, to estimate repair times. 
Remarkably, fault injection techniques turned out to be useful when empirical data are lacking, and revealed that time-to-recovery is much longer than what is typically assumed by most model-based studies.

\vspace{3pt}
    
The rest of the paper is organized as follows: Section~$2$ discusses relevant work on availability assessment in cloud environments. In Section~$3$ we provide an overview of the IMS case study, with a discussion about its containerized deployment. In Section~$4$ we present the availability and queueing models of a CNF, being the elementary structure of a service chain. In Section~$5$ we address the chain availability concern through the MUGF technique and the related optimization problem. Section~$6$ presents the testbed and offers details about the experimental trials. Section~$7$ concludes the paper. 
For the sake of readability, Table~\ref{tab:notation} summarizes the notation adopted across the paper.

\section{Related work}

Availability issues represent a hot topic when dealing with {\em softwarized} networks, where the presence of virtualized entities (e.g. hypervisors, containerized environments, etc.) pose new intriguing challenges for telecom operators that are called for adhering to strict Service Level Agreements (SLAs) \cite{tola1,tola2,tola3,qu17}. Due to the vastness of the topic, technical studies adopt different angles to face the availability problems, including: designing available virtualized infrastructures to manage traffic problems \cite{traffic1,traffic2}, optimizing the allocation of virtualized infrastructures to maximize the resiliency \cite{alloc1,alloc2}, optimizing the availability scheduling of virtual resources \cite{schedule1, schedule2}, managing the state of virtualized services in resiliency problems \cite{state1, state2}. 
On the other hand, in our work we mainly focus on a modeling methodology for the availability issues in softwarized networks. Thus, we are going to explore some affine literature more in detail, by highlighting the main differences with our work.

	\begin{table}[t!] 
		\caption {Notation} \label{tab:notation}
		\centering
		\resizebox{.5\textwidth}{!}
		{
			\renewcommand{\arraystretch}{1.5}
			\begin{tabular}{c|c}
				\hline
				$m$ & (Containerized) IMS tier (P,S,I,H) \\
				$\ell$ & CNF redundancy index \\
				CNF$^{(m,\ell)}$ &  Parallel CNF $\ell$ associated to tier $m$ \\
				$i$; $k$ & Tenant $i$; number of tenants using the cIMS \\
				$\eta_i$ & Number of working containerized instances controlled by tenant $i$ \\
				\boldmath$\eta$ & CNF State vector \\
				$N^{(m,\ell)}$ & Number of states of CTMC performance model of CNF $\ell$ of tier $m$  \\
				$g_{i, \boldsymbol{\eta} }$ & Capacity level exposed by CNF for tenant $i$ in state {\boldmath$\eta$} \\
				$\gamma$ & Serving capacity \\
				$\bm{g}_{\boldsymbol{\eta}}$ &  Capacity level vector in state $\boldsymbol{\eta}$ \\
				$\bm{\delta}_{\boldsymbol{\eta}}$ &  Mean delay performance vector in state $\boldsymbol{\eta}$ \\
				$p_{\boldsymbol{\eta}}$  & Steady-state probability of being in state $\boldsymbol{\eta}$ \\ 
				$\boldsymbol{\Delta}(t)$ & Vector stochastic process including all tenants mean delays per CNF \\
				$\boldsymbol{W}^c(t)$ & Maximum tolerated value for the mean CSD \\
				$A^c(\bm{w}^c)$ & Steady-state availability of the cIMS \\
				$\bm{J}^c$ & Number of states of the cIMS \\
				$u^{(m)}(\bm{z}) ; u^{c}(\bm{z})$ & MUGF for tier $m$; MUGF for the cIMS  \\
				$\Omega_S, \Omega_D, \Omega_I$ & State spaces of software, docker, infrastructure layers \\
				$\lambda_C, \lambda_D, \lambda_I$ & Failure rate of containerized instances, docker, and infrastructure layers \\
				$\mu_C, \mu_D, \mu_I$ & Repair rate of containerized instances, docker, and infrastructure layers \\
				$\alpha_i$ & Arrival rate of requests at tenant $i$ \\
				$\beta_m$ & Service rate of requests at tier $m$ \\
				\hline
				\hline
			\end{tabular}
		}
	\end{table}

Fan et al. \cite{fan2017} faced an availability problem concerning the optimal deployment of an SFC infrastructure. In particular, their aim is to find the minimum number of backup VNFs that guarantees a desired availability level. 
A similar problem is tackled by Kong et al. \cite{kong2017}, where a heuristic algorithm has been conceived to maximize the availability of an SFC through an optimal distribution of backup VNFs across primary and backup paths. Alameddine et al. \cite{alameddine2017} focused on virtual machine redundancy in a multi-tenant environment by adopting an optimal primary/backup logic. All of these works did not consider, or only partially considered, a failure/repair model that is instead accurately examined in our availability setting.

Other studies focused on more compact formalisms to model network availability aspects.
It is the case of Sebastio et al. \cite{sebastio18}, where an availability assessment of exemplary containerized architectures is faced through the Stochastic Reward Networks (SRNs) framework. An SRN-based approach has been profitably adopted also by Bruneo \cite{bruneo14}, where a stochastic model to typify some aspects of an Infrastructure-as-a-Service framework has been considered. Similarly, a technique relying on Stochastic Petri Networks (SPNs) has been exploited by Sousa et al. \cite{sousa14}, where an availability analysis of cloud-based deployments has been carried out. 
Interestingly, both SRN/SPN and the proposed MUGF rely on a common underlying Markov model. While MUGF prefers an open analytical approach, SRN/SPN offers a more compact representation (through the formalism of places, arcs, tokens) that can be more convenient for some users. 
However, in the case of multi-tier systems such as SFCs, as further discussed in this paper, having the underlying Markov model hidden by the SPN/SRN hinders the computation of the availability.

Another track of works exploits the UGF methodology, which, although historically adopted to cope with availability issues in the field of industrial systems, has found fertile ground in networking management. Some examples include the work by Sun et al. \cite{Sun2016}, where a UGF-based technique is assessed for modeling physical and virtual system failures, and Yu et al. \cite{Yu2015}, where the UGF has been applied in the field of service requests in cloud scenarios. Both studies focused on single-tenant environments, not calling for the application of a multidimensional form of UGF.  

In the present paper, by exploiting the properties of the multidimensional UGF (MUGF) at first conceived by our previous work \cite{MDMTSC}, we propose a new availability assessment method for multi-tenant environments. 
A set of clear novelties emerge with respect to the original proposal. First, in this work we consider a containerized infrastructure (in place of a traditional virtualized architecture considered in previous work \cite{MDMTSC}), which is reflected in the three-layered structure of our model, where the containers do not embed an OS and are deployed on top of the Docker container manager. This model directly translates into a novel experimental testbed (missing in the previous work \cite{MDMTSC}), based on the Clearwater project, a realistic cIMS deployment which allows to estimate the value of key quantities such as repair rates, call setup latencies, and mean service times. Moreover, in this work each tenant is modeled in accordance to a sophisticated queueing model (there was no queueing model in the previous work \cite{MDMTSC}), which allows to analyze the \emph{call setup delay} (CSD), a critical latency metric for $5$G networks as specified by ETSI \cite{etsicsd}. Finally, the MUGF structure (in particular, series and parallel operators) is totally different from the one introduced in previous work due to the different metric adopted: the number of sessions in \cite{MDMTSC}, and the CSD in this new proposal.

We remark that, from a scalability viewpoint, the MUGF technique exhibits interesting results compared to other methods in similar fields. For example, Petri net structures (e.g., SPN/SRN), beyond requiring very specific tools to be solved, can suffer when dealing with large and nested systems, as also highlighted by Peterson \cite{petri1} and Herzog \cite{petri2}. 
In particular, SRNs offer a system state representation in terms of the ``token'' distributions (a.k.a. markings), where each marking is representative of a particular state of the system at a given time $t$. 
On one hand, this approach provides a comfortable interface to automatically specify the token distribution by hiding technical details about the underlying state model. On the other hand, such an approach does not allow to easily retrieve the MSS state distribution which is needed by the analytical formulation of the MUGF. 

Yet, the classic Continuous-Time Markov Chain (CTMC) representation would lead to a space state explosion, since it requires the entire cIMS to be modeled (monolithic approach), whereas the MUGF uses a combination of performance distributions of single nodes to achieve the cIMS performance distribution (decomposition approach).

\section{Containerized IMS case study}
\label{sec:ims}

\begin{figure}[t]
	\centering
	\captionsetup{justification=centering}
	\includegraphics[scale=0.22,angle=90]{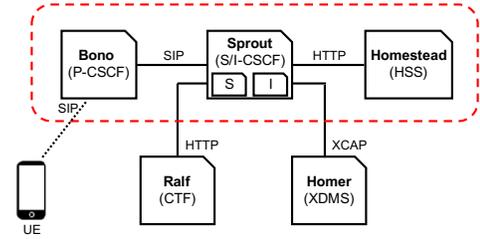}
	\caption{Overview of the Clearwater IMS.}
	\label{fig:ims}
\end{figure} 

\begin{figure}[t]
	\centering
	\captionsetup{justification=centering}
	\includegraphics[scale=0.28,angle=90]{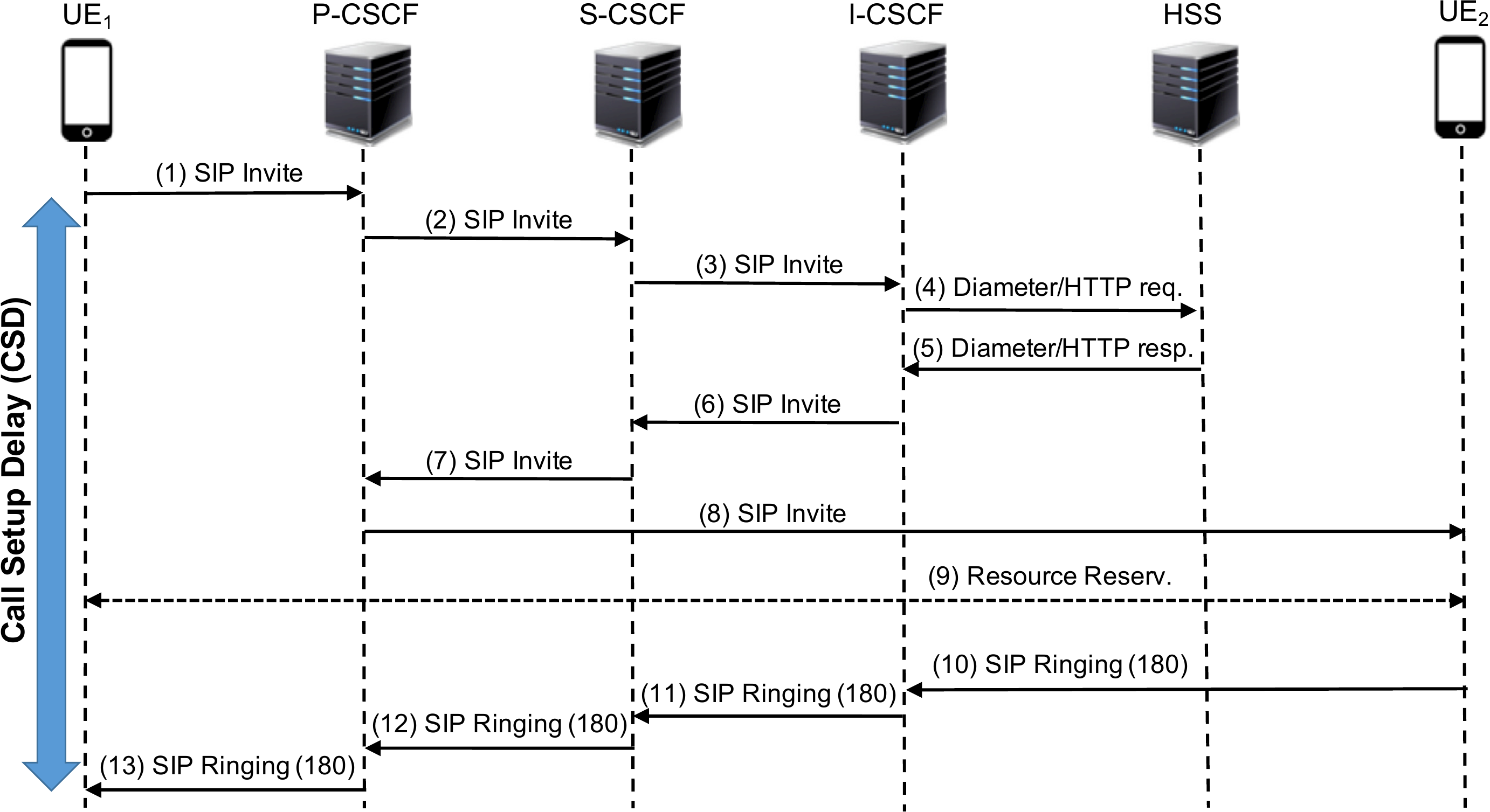}
	\caption{IMS (single domain) call setup, where the Call Setup Delay metric is represented on the left.}
	\label{fig:csd}
\end{figure} 

In this study we present the proposed approach in the context of a cIMS case study, based on Clearwater \cite{clearwater}, a real IMS product that fully leverages containers and cloud computing technology. 
The cIMS consists of a chain of \emph{softwarized} functions running within containers. Such containers are managed by a container engine (in our case Docker) which is installed on a physical machine (a \emph{node}). In turn, the nodes can be replicated to form a \emph{tier}, in order to achieve higher performance and availability. Figure \ref{fig:ims} shows a sketch of the Clearwater IMS characterized by the following functions:
\textit{Bono}, which is the P-CSCF (Proxy-Call Session Control Function), and acts as anchor point for clients relying on the Session Initiation Protocol (SIP); 
\textit{Sprout} simultaneously acts as S-CSCF (Serving-CSCF) in charge of managing SIP registrations, and as I-CSCF (Interrogating-CSCF) for handling associations between UEs and a specific S-CSCF; 
\textit{Homestead} represents the HSS (Home Subscriber Server) for users authentication; 
\textit{Ralf} acts as CTF (Charging Trigger Function), for charging and billing operations; 
\textit{Homer} manages service setting documents per user, acting as XML Document Management Server (XDMS). 
A red dashed rectangle in Fig. \ref{fig:ims} indicates the mandatory functions of the cIMS architecture that we consider in our analysis. 
 
The IMS is called to satisfy real-time constraints such as delay, jitter, packet loss. In this regard, a metric called Call Setup Delay (CSD) has been elected as a critical Key Performance Indicator (KPI) \cite{csd1,csd2,csd3}, being strongly related to the end-user experience. Formally, CSD is a time-based metric defined as the time interval between Invite message sent from the caller and the received Ringing message (code $180$) \cite{begain-book}, and well approximates the average time that user requests spend in the cIMS (due to the processing time needed by each node to handle the requests), where the propagation delays are neglected. 
Figure \ref{fig:csd} shows a simplified scenario of a SIP call flow (IMS single domain), where the CSD is accordingly represented as a vertical double arrow between the initial sent message, (1) SIP Invite, and the last received message, (13) SIP Ringing.

\subsection{Containerized Network Function Model}
\label{subsec:contims}

The network functions of a service chain can be deployed in dedicated containers, and decoupled from the underlying infrastructure, according to a three-layer structure: 

\begin{itemize}

	\item {\em Software layer}: Its role is to run application software that implements the business logic of the service chain, to be deployed as containers (for example, in our testbed, the \emph{Bono} and the \emph{Sprout} applications). We assume that a CNF hosts containers of the same type;
	
	\item {\em Docker layer}: Its role is to provide a run-time support (e.g., a service daemon) to build, run, and manage OS containers; this layer is also exploited in other container management technologies, such as \emph{Linux Container Daemon} and \emph{Rocket};
	
	\item {\em Infrastructure layer}: It represents the underlying physical layer that, for the sake of simplicity, includes only the operating system (OS) and the hardware (HW) components (e.g., CPU, RAM, etc.).

\begin{figure}[t!]
	\centering
	\captionsetup{justification=centering}
	\includegraphics[scale=0.3, angle=90]{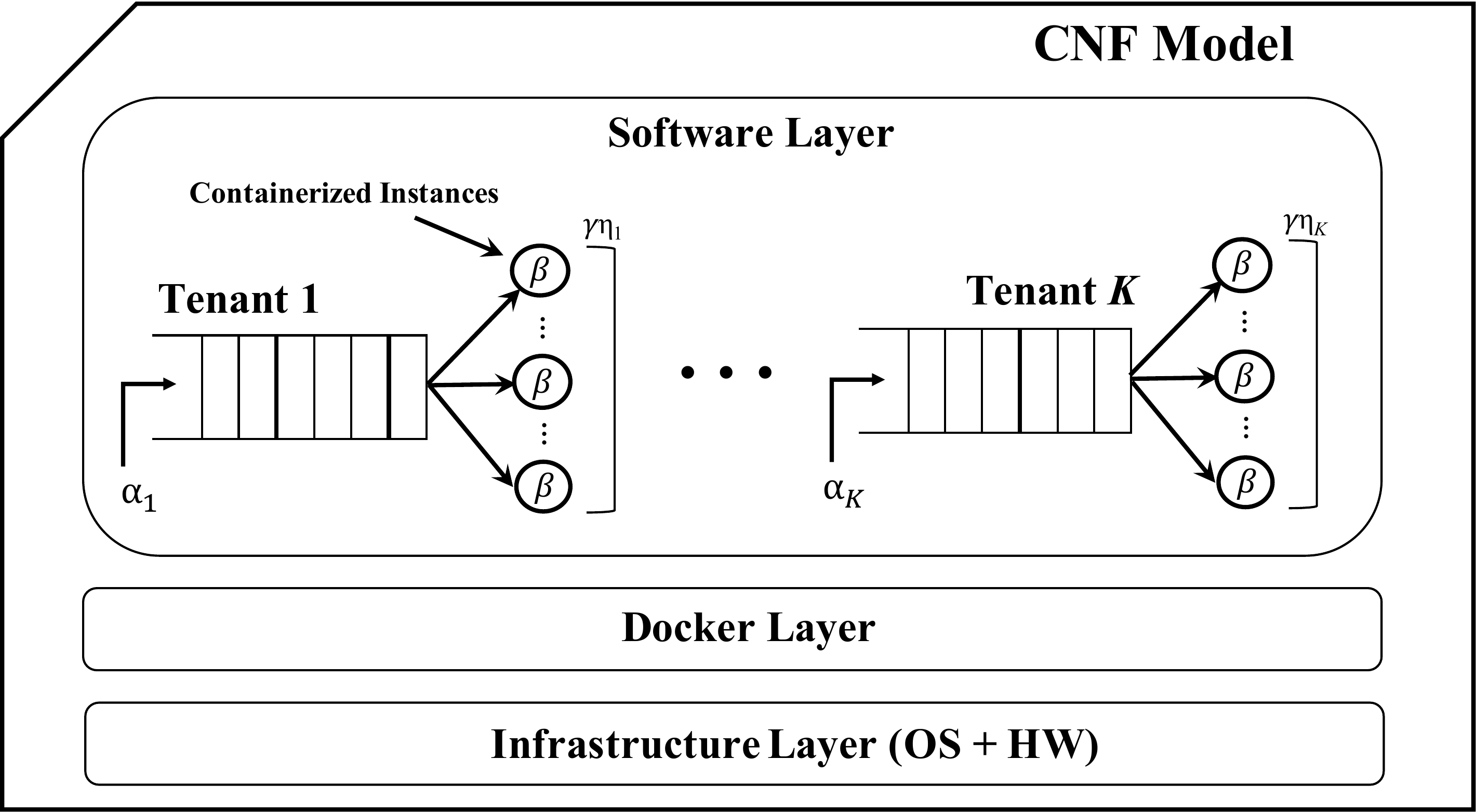}
	\caption{A CNF hosting $K$ tenants. Each tenant can be represented through a $M/G/\gamma \eta_i$ queueing model managing a set of containerized instances in the Software layer. Quantities $\alpha_i$ and $\beta$ are the arrival and service rates, respectively. The Docker layer manages containers. The Infrastructure layer embodies the host OS and the hardware.}
	\label{fig:cnf_model}
\end{figure} 	
	
\end{itemize}

\begin{figure*}[t!]
	\centering
	\captionsetup{justification=centering}
	\includegraphics[scale=0.55, angle=90]{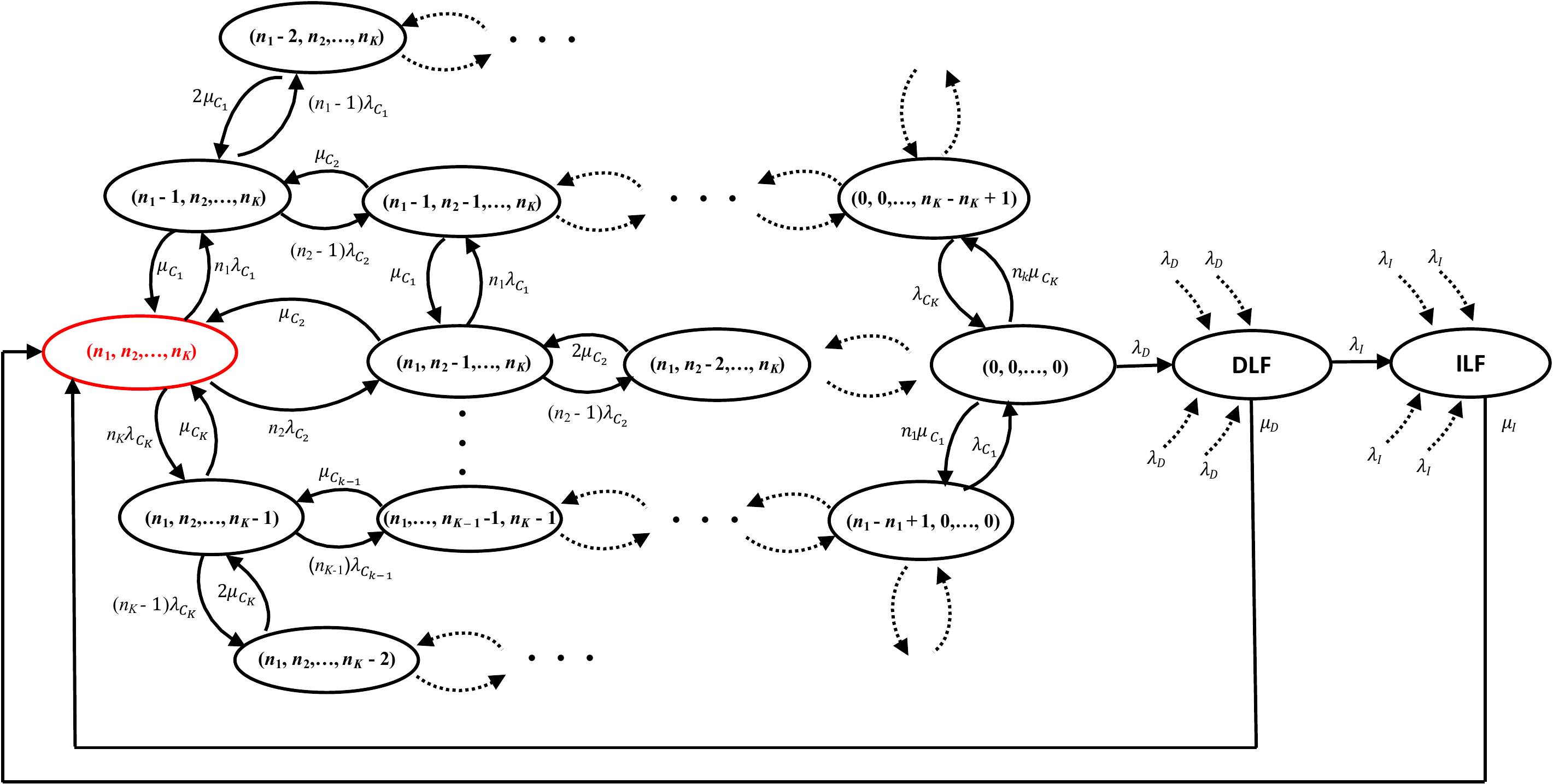}
	\caption{Transition-state diagram of CNF multi-state model. States are labeled with a vector that represents, for each tenant $1 \dots K$, how many container instances are available. The state vector $(n_1,...,n_K)$ in red indicates a fully-working system that runs the maximum number of container instances per tenant. States DLF and ILF indicate Docker and Infrastructure layers failures, respectively.} 
	\label{fig:sephi1}
\end{figure*} 
\noindent A single CNF represents the elementary block of each tier of the service chain (i.e., a node in the system that includes the three layers). As it will be clear in the following, a cIMS tier can be made of several redundant CNFs (typically, of one and the same type), in order to increase capacity and to meet performance and availability objectives. A stylized representation of a CNF is depicted in Fig. \ref{fig:cnf_model}, where the $i$-th tenant ($i=1,\dots,K$) manages $\eta_i$ containerized instances. 
For each tenant, a containerized instance can manage, in turn, a number of service requests amounting to $\gamma$ (\textit{serving capacity}), at the same time. In practice, a containerized instance is supposed to be composed of processes, each one handling a single request. 
The serving capacity represents the number of requests handled by each tenant. 
The resulting queueing model of a single CNF for tenant $i$ is an infinite queue $M/G/\gamma \eta_i$ as in Fig. \ref{fig:cnf_model}, where $\gamma \eta_i$ depends on the actual working conditions of CNF (see the forthcoming Sect. \ref{sect:queuemodel} for more details). 
Moreover, all tenants share the Docker and Infrastructure layers, as typical of modern cloud deployments. 
Note that Fig. \ref{fig:cnf_model} represents the system from a conceptual perspective, as in its actual implementations the load among container instances is balanced by communication mechanisms (such as message queues, or DNS-based load balancing) managed by the underlying Docker and infrastructure layers.

\section{CNF Modeling}
\label{sect:avamodel}
In this section, we analyze the behavior of a single CNF according to a double perspective. The first one concerns the availability characterization of a CNF in terms of a multi-state model, where the failure/repair behaviors of the three layers (Software, Docker, and Infrastructure) are represented and evaluated (Sect. \ref{sect:avacnf}). The second one pertains to the latency characterization by means of CNF queueing modeling (Sect. \ref{sect:queuemodel}). 
The CNF availability model and the CNF queueing model can both be applied on a given CNF configuration, in order to get an assessment of its availability and latency. These results will be used in the next section for capacity planning of the service chain.

\subsection{CNF availability model}
\label{sect:avacnf}

The stochastic behavior (in terms of failure and repair actions) of the three CNF layers can be captured through the concept of {\em state}. A state reflects a particular condition (e.g. up/down) of: one or more containerized instances (belonging to the Software layer), the Docker layer, the Infrastructure layer. Accordingly, the CNF model of Fig. \ref{fig:cnf_model} can be {\em translated} in terms of the transition-state diagram reported in Fig. \ref{fig:sephi1}, where a state is labeled with a $K$-dimensional vector {\boldmath$\eta$} = $(\eta_{1},...,\eta_{K})$ $\in \prod_{i=1}^{K}\{0,\dots,n_i\}$ and $\eta_{i}\in\{0,1,...,n_i\}$ represents the number of working container instances that belong to tenant $i$. The initial state vector $(n_1,...,n_K)$ represents a fully-working system that runs the maximum number of container instances per tenant (reported in red in Fig. \ref{fig:sephi1}). 
As faults occur, the system moves to states with lower values in the vector. In the worst case, all container instances are failed (all values in the vector are zero). Similarly, as container instances are recovered, the system moves to states with higher values in the vector. 
For instance, the vector $(n_1,...,n_i - 1,...,n_K)$ indicates a state where one-out-of-$n_i$ container instances of the $i$-th tenant is failed. All the failure interarrivals are supposed to be independent and identically distributed (\textit{iid}) random variables according to an exponential distribution with parameter $\lambda_{C_i}$. The duration of a repair action of a failed containerized instance is assumed to be an exponential random variable with parameter $\mu_{C_i}$. All the failure and repair times are supposed to be independent of each other. Thus, the transition rates are proportional to the number of container instances that can fail and that can be recovered at each state.
The state \textit{Docker Layer Failure} (DLF) indicates the Docker failure condition which, in turn, causes the failure of all the containerized instances, and the corresponding state vector is $(0,...,0)_D$. 
For Docker too, failure interarrivals are supposed to be \textit{iid}  according to an exponential distribution with parameter $\lambda_{D}$, and independent from containerized instances failures. The duration of a repair action of this layer is assumed to be an exponential random variable with parameter $\mu_{D}$. When Docker restarts, all the containerized instances are assumed to be restarted. Such a condition is taken into account by the transition from DLF state to initial (completely working system) state with rate $\mu_{D}$.

	The state \textit{Infrastructure Layer Failure} (ILF) is associated with a crash of HW/OS part provoking a failure of the entire system. Again, in this case the state vector is $(0,...,0)_I$. Also for the Infrastructure layer, failure interarrivals are supposed to be \textit{iid} according to an exponential distribution with parameter $\lambda_{I}$, and independent from upper layers failures. The duration of a repair action of this layer is assumed to be an exponential random variable with parameter $\mu_{I}$.  Similar to the previous case, when Infrastructure is restored, both Docker and Software layers are restored. Such a condition is taken into account by the transition from ILF state to initial (completely working system) state with rate $\mu_{I}$.

		

It is useful to highlight that, the assumptions on the considered repair rates stem from the fact that mean-time-to-repairs (namely $1/\mu$) are approximately constant over time. It is especially valid for the software infrastructures (as in our case) where repair actions are meant to be reboot actions, as also confirmed by credited literature \cite{trivedi-book, trivedi-perfeval, trivedi-bobbio}.

To derive the steady-state availability, we formally model the CNF as a Multi-State System (MSS). Let $\Omega_S=\prod_{i=1}^{K}\{0,1,\dots,n_i\}$ be the state space of the Software layer, being $\{0,1,\dots,n_i\}$ the state space of the tenant $i$, $i=1,\dots,K$; let $\Omega_D=\{0,1\}_D$ and $\Omega_I=\{0,1\}_I$ be the state spaces of Docker and Infrastructure layers, respectively, where $0$ indicates the failure condition whereas $1$ refers to the working condition. Thus, the state space of the overall CNF is $\Omega=\Omega_S \times \Omega_D \times \Omega_I$.
	
Moreover, we define the \textit{capacity level} $g_{i, \boldsymbol{\eta} }$ exposed by the CNF for tenant $i$ in state $\boldsymbol{\eta}$ as
\begin{equation}
g_{i,\boldsymbol{\eta}} = \gamma \cdot \eta_{i},
\label{eq:gperf}
\end{equation} 
where $\gamma$ has been previously defined as the serving capacity of a containerized instance for each tenant.
Since the capacity of CNF can vary over time due to failures, the capacity level $g_{i,\boldsymbol{\eta}}$ is time-varying.
The set including all possible capacity levels $\bm{g}_{\boldsymbol{\eta}}=(\bm{g}_{1,\boldsymbol{\eta}},...,\bm{g}_{K,\boldsymbol{\eta}})$ for each CNF is
\beq
\mathcal{G} =  \left\{ \gamma \cdot \boldsymbol{\eta} \Biggm| \boldsymbol{\eta} \in \prod_{i=1}^{K} \{0,\dots,n_i\} \right\} \cup \{ {(0,\dots,0)_D}, {(0,\dots,0)_I}\},
\label{eq:gvec}
\eeq
where ${(0,\dots,0)_D}$ and ${(0,\dots,0)_I}$ refer to the capacity levels of the DLF and ILF states, respectively.
The total number of states in Fig. $5$ is
\begin{equation}
N = |\Omega|= \prod_{i = 1}^K\left(n_i + 1\right) + 2.
\label{eq:numstati}
\end{equation}

To characterize the considered MSS, it is useful to introduce the \textit{structure function} $\varphi: \Omega \rightarrow   \left\{ \boldsymbol{\eta} \Biggm| \boldsymbol{\eta} \in \prod_{i=1}^{K} \{0,\dots,n_i\} \right\} \cup \{(0,\dots,0)_D,(0,\dots,0)_I \}  $ defined as follows:
	\begin{align}
	\centering
	\varphi( \mbox{\boldmath$\eta$}, x_D,x_I) = 
	\left\{
	\begin{array}{l}
	\mbox{\boldmath$\eta$}, ~~~~~~~~~~~ x_D=1, x_I=1 \\ 
	(0,\dots,0)_D, ~~ x_D=0, x_I=1 \\
		(0,\dots,0)_I, ~~ x_I=0,
	\end{array}
	\right.
	\label{strfung}
	\end{align}
where $x_D \in \Omega_D$ and $x_I \in \Omega_I$.
It is a deterministic function \cite{Levibook} useful to specify the relation between the states of single elements (layers) and the state of the overall system (CNF).
Now, the MSS of the CNF is completely described by the state space $\Omega$, the serving capacity levels $\left\{ \boldsymbol{\eta} \Biggm| \boldsymbol{\eta} \in \prod_{i=1}^{K} \{0,\dots,n_i\} \right\} \cup \{(0,\dots,0)_D,(0,\dots,0)_I \}$, and the structure function $\varphi$.  
Let now be $\boldsymbol{X}(t)=(X_1(t),\dots,X_K(t))$ a $\Omega_S$-valued stochastic process which denotes the failure-repair process of the Software layer; $X_D(t)$ a $\Omega_D$-valued stochastic process which denotes the failure-repair process of Docker layer; $X_I(t)$ a $\Omega_I$-valued stochastic process which denotes the failure-repair process of Infrastructure layer, all defined for $t \geq 0$. Each of the above processes is Markovian on its state space, being all the underlying random variables exponentially distributed and independent of each other. 
Thus, the stochastic process $\{ \varphi(\boldsymbol{X}(t), X_D(t), X_I(t)), t \geq0 \}$ is represented by the transition-state diagram in Fig. $5$. 
Moreover, the CNF capacity level at time $t\geq0$ is expressed in terms of the vector stochastic process 
\begin{equation}
\bm{G}(t) = \left(G_1(t),...,G_K(t)\right) = \gamma \cdot \varphi(\boldsymbol{X}(t), X_D(t), X_I(t)) , 
\label{eq:gvecstoc}
\end{equation}  
with values in $\mathcal{G}$, and with (state) probability vector $\bm{p}(t)$ at time $t$ collecting all the state probabilities $p_{\boldsymbol{\eta}}(t)$, being $p_{\boldsymbol{\eta}}(t) = \mathrm{Pr}\{\bm{G}(t) = \bm{g}_{\boldsymbol{\eta}}\}$.
We note that $\bm{G}(t)$ is a CTMC process described by a transition-state diagram equal to the one in Fig. $5$, except that each state vector must be multiplied by $\gamma$.

We see that $\bm{G}(t)$: $i)$ has a finite state space with cardinality $N$ given by (\ref{eq:numstati});  $ii)$ is irreducible, since every state is reachable from every other state (see Fig. $5$); $iii)$ is homogeneous since its infinitesimal generator matrix $\mathbf{Q}$ has constant elements (given constant parameters of failure and repair random variables). 
From the above properties, $\bm{G}(t)$ is an ergodic CTMC with a unique steady-state probability vector $\bm{p}=\lim_{t\rightarrow\infty} \bm{p}(t)$, with $\bm{p}(t)$ given by the solution of 
\begin{equation}
\frac{{\rm d}\bm{p}(t)}{{\rm d}t} = \bm{p}(t) \mathbf{Q},
\label{eq:diff_eq}
\end{equation} 
 along with the normalization condition $\sum_{\boldsymbol{\eta}} p_{\boldsymbol{\eta}}(t)= 1$.
 
Again, $\bm{p}$ is a vector collecting the state-state probability $p_{\boldsymbol{\eta}}$ for each state $\boldsymbol{\eta}$, that is given by:

\begin{equation}
p_{\boldsymbol{\eta}} = \lim_{t\rightarrow\infty} p_{\boldsymbol{\eta}}(t) = \lim_{t\rightarrow\infty} \mathrm{Pr}\{\bm{G}(t) = \bm{g}_{\boldsymbol{\eta}}\}.
\label{eq:pi}
\end{equation}
Accordingly, the (discrete) random vector $\bm{G} = \left(G_1,...,G_K\right) \in\mathcal{G}$ corresponds to the asymptotic behavior of $\bm{G}(t)$ (in the limit of $t\rightarrow\infty$) and admits values in the set (\ref{eq:gvec}) with probabilities (\ref{eq:pi}). 
In conclusion, the set of pairs $\left\{p_{\boldsymbol{\eta}}, \bm{g}_{\boldsymbol{\eta}} \right\}$ determines the steady-state behavior of a CNF in terms of serving capacity.\\

\subsection{CNF Queueing model}
\label{sect:queuemodel}


Our queueing model reflects the typical behavior of real systems, where the requests arriving to the CNF are served according to a queue characterized by non-exponential service times. 
As regards the choice of the queueing model system, we lie not so far from technical literature where SIP-based servers are characterized through $M/M/1$ or $M/M/k$ infinite queueing systems (see \cite{shulz,Subramanian_2009,Subramanian_2011,dimauro19,Guida2013}) with a fixed number of servers. 
Such models are based on the exponential assumption of service times that, in the field of network communications, might be quite unrealistic.
In our work, indeed, we are able to directly measure the service times across our testbed (see Sect. \ref{sec:exp}.1), and we find out that the empirical distribution of service times is not well fitted by an exponential distribution. Accordingly, we adopt an $M/G/G_i(t)$ model accounting also for the time-varying number of servers due to the failure/repair process for each tenant $i$. 
For a given state of the model, $G_i(t)=\gamma \eta_i$ and thus $M/G/G_i(t)$ is equivalent to $M/G/\gamma \eta_i$.


Furthermore, we want to highlight an important relationship between queueing and failure/repair models in terms of time scales. As observed in \cite{whittbook}, in the field of communication networks it is possible to distinguish various and different time scales, such as the failure time scales (FTS) and the service time scales (STS). The former governs the failure/repair processes, and the latter governs the typical queueing metrics (e.g. the service times). When a time scale completely dominates another one, it is possible to neglect the transient effects produced by the dominated time scale. 
In service chains such as the cIMS, FTS $\gg$ STS, since FTS is in the order of thousands of hours, while STS is in the order of milliseconds (see also the parameters in Table \ref{tab:params}). Thus, a decoupling between FTS and STS can be reasonably assumed. This behavior leads to claim that, given a state $\boldsymbol{\eta}$, tenant queues reach their steady-states very quickly compared to the occurrence of faults.
In summary, for each state $\boldsymbol{\eta}$, we use a steady-state $M/G/\gamma \eta_i$ queue model for tenant $i$ to derive the delay model of a CNF. 
	

We assume that call setup requests arriving to tenant $i$ follow an arrival Poisson process (a common assumption in technical literature \cite{pois1,pois2,pois3}) with parameter $\alpha_i$. Let $1/\beta$ be the mean value of service times derived from the experiments, that we suppose to be one and the same for all requests and tenants.
Now, since no closed forms are available to evaluate the mean delay introduced by the $M/G/\gamma \eta_i$ model to requests, we proceed in two steps: $i)$ we derive the mean delay for $M/M/\gamma \eta_i$; $ii)$ then, we compute the mean delay for the $M/G/\gamma \eta_i$ model by using the approximating formula known as the Kingman's law of congestion (see \cite{kimura,kingman_tech}), that exploits the coefficients of variation of the measured service times. 
	
Along with the first step, according to \cite{kleinbook}, the mean number of requests (or jobs) $\nu_{i,\boldsymbol{\eta}}$ for the $M/M/\gamma \eta_i$ model for tenant $i$ in state $\boldsymbol{\eta}$ is:

\begin{equation}
E[\nu_{i, \boldsymbol{\eta}}]=\gamma \eta_i\cdot \rho_{i, \boldsymbol{\eta} } + \rho_{i, \boldsymbol{\eta} } \frac{(\gamma \eta_i \cdot \rho_{i, \boldsymbol{\eta} })^{\gamma \eta_i}}{\gamma \eta_i !} \frac{\pi_{i, \boldsymbol{\eta} }}{(1-\rho_{i, \boldsymbol{\eta} })^2},
\label{eq:sess}
\end{equation}
where {\em utilization factor} amounts to:
\begin{equation}
\rho_{i, \boldsymbol{\eta} }=\frac{\alpha_i}{\beta \cdot \gamma \eta_i}, \nonumber
\end{equation}
and 
\begin{equation}
\pi_{i, \boldsymbol{\eta} }=\left[ \sum_{h=0}^{\gamma \eta_i -1 } \frac{(\gamma \eta_i \cdot \rho_{i, \boldsymbol{\eta} })^h}{h!} + \frac{(\gamma \eta_i \cdot \rho_{i, \boldsymbol{\eta} })^{\gamma \eta_i}}{\gamma \eta_i !} \frac{1}{1-\rho_{i, \boldsymbol{\eta} }}  \right]^{-1}.  \nonumber
\end{equation}
By virtue of Little's law, the mean delay is
\begin{equation}
 E[d_{i, \boldsymbol{\eta} }]=\frac{E[\nu_{i, \boldsymbol{\eta} }]}{\alpha_i}.
\label{eq:delay}
\end{equation} 
In the second step, since the first and second moments of service times distribution are finite, we apply the Kingman's approximation to derive the mean delay introduced by the considered $M/G/\gamma \eta_i$ system (say $\delta_{i, \boldsymbol{\eta} }$) of tenant $i$ in state $\boldsymbol{\eta}$, namely
\begin{equation}
\delta_{i, \boldsymbol{\eta} } \approx E[d_{i, \boldsymbol{\eta} }] \cdot \frac{1+CV_s^2}{2},
\label{eq:kingman}
\end{equation} 	
being $CV_s$ the coefficient of variation of the empirical service time\footnote{Such an approximation holds either for individual queues and for open non-Markovian network of queues (see \cite{kimura,whitt_net_queue}).}. 
As pointed by Whitt (see \cite{whitt}), such an excellent approximation can be considered a special case of Allen-Cunneen approximation \cite{allenbook}. Obviously, $\delta_{i, \boldsymbol{\eta} }$ increases as $\gamma \eta_i$ reduces due to failures. On the other hand, $\delta_{i, \boldsymbol{\eta} }$ decreases as $\gamma \eta_i$ increases by virtue of repair actions.
It is worth noting that, the Kingman's approximation can be easily generalized also to the case of generic arrival times, with a little modification of eq. (\ref{eq:kingman}) which can be rewritten as: $\delta_{i, \boldsymbol{\eta} } \approx E[d_{i, \boldsymbol{\eta} }] \cdot (CV_a^2+CV_s^2)/2$, being $CV_a^2$ the coefficient of variation of the distribution of the arrivals. 

Being $\mathcal{B}^K$ the Cartesian product of a set $\mathcal{B}$ for itself $K$ times, to characterize the MSS describing the delay model, it is useful to introduce the structure function  $\varphi_{\Delta}: \Omega \rightarrow \{\mathbb{R}^+ \cup \{+\infty\} \}^K$ defined as follows:
\begin{align}
\centering  
\varphi_{\Delta}( \boldsymbol{\eta}, x_D,x_I) = 
\left\{
\begin{array}{l}
\left(\delta_{1, \boldsymbol{\eta} },\dots,\delta_{K, \boldsymbol{\eta} } \right), ~~x_D=1, x_I=1 \\ 
(+\infty,\dots,+\infty), ~~~~ x_D=0, x_I=1 \\
(+\infty,\dots,+\infty), ~~~~ x_I=0,
\end{array}
\right.
\label{strfundelta}
\end{align}	
where $\delta_{i, \boldsymbol{\eta} }$ is given by (\ref{eq:kingman}), $i=1,\dots,K$, and where the infinite delay arises when there are no working containerized instances. The mean delays introduced by a single CNF at time $t \geq 0$ is the vector stochastic process 
\begin{equation}
\bm{\Delta}(t) = \left(\Delta_1(t),...,\Delta_K(t)\right) = \varphi_{\Delta}(\boldsymbol{X}(t), X_D(t), X_I(t)).
\label{eq:delvecstoc}
\end{equation}  
	
Similarly to the stochastic process $\bm{G}(t)$ in (\ref{eq:gvecstoc}), $\bm{\Delta}(t)$ is an ergodic CTMC, and the random vector $\bm{\Delta} = \left(\Delta_1,\dots,\Delta_K\right)$ corresponds to the asymptotic behavior of $\bm{\Delta}(t)$ as $t \rightarrow \infty$. 
Given $\bm{\delta}_{\boldsymbol{\eta} } = \left(\delta_{1, \boldsymbol{\eta} },...,\delta_{K, \boldsymbol{\eta} } \right)$ the mean delays vector for each state $\boldsymbol{\eta}$, the set of pairs $\left\{p_{\boldsymbol{\eta} }, \bm{\delta}_{\boldsymbol{\eta} } \right\}$ determines exhaustively the steady-state performance behavior of the CNF in terms of mean delay, being $p_{\boldsymbol{\eta}}$ given by (\ref{eq:pi}).





\section{Modeling of the CNF service chain}
\label{sect:ava}

\begin{figure}[t!]
	\centering
	\captionsetup{justification=centering}
	\includegraphics[scale=0.3,angle=90]{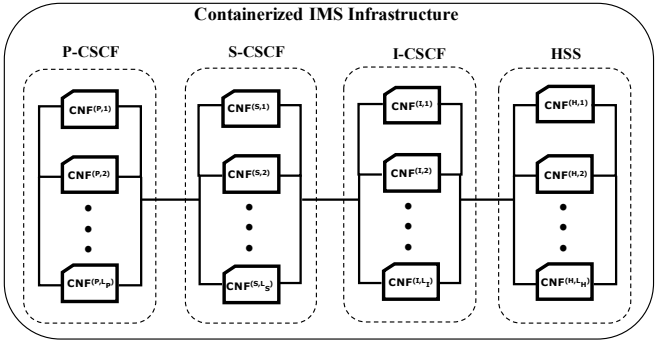}
	\caption{Multi-tenant cIMS system, where parallel CNFs are introduced for redundancy purposes. CNF$^{(m,\ell)}$ refers to parallel CNF $\ell$ ($\ell=1,\dots,L_m$) of tier $m$, with $m\in$\{P-CSCF, S-CSCF, I-CSCF, HSS\}.}
	\label{fig:serpar}
\end{figure}

Based on the CNF modeling from the previous section (which represents an individual CNF in the service chain), we here build a model for the whole service chain of multiple CNFs. Two important aspects emerge. First, the chain is made of tiers {\em connected in series} (e.g., see Fig. \ref{fig:serpar} in the context of the cIMS). Thus, the entire chain is supposed to be working when every tier $m$ in the chain is working (e.g., $m\in $ \{P, S, I, H\} for the cIMS, where P, S, I, H, indicate for brevity P-CSCF, S-CSCF, I-CSCF, HSS, respectively). Second, each tier $m$ consists of redundant CNFs {\em connected in parallel}, in order to improve performance and availability. We remark that a tier $m$ acts as a {\em logical} entity, by dividing the load among the replicas in the tier, according to the \emph{flow dispersion} hypothesis (any CNF is able to handle service requests - see \cite{Levitin2003}). Every replica includes all of the three layers of the CNF structure (Software, Docker, and Infrastructure).

We denote with CNF$^{(m,\ell)}$ the $\ell$-th parallel CNF associated to tier $m$ ($\ell=1,\dots,L_m$). 

Now, we start to evaluate the mean CSD introduced by tier $m$, where each CNF$^{(m,\ell)}$ is modeled as an $M/G/G_i^{(m,\ell)}(t)$ queue for tenant $i$. Indeed, tier $m$ is given by $L_m$ parallel CNFs (see Fig. \ref{fig:serpar}), and can be analyzed as a single $M/G/G_i^{(m)}(t)$ queue as consequence of the flow dispersion hypothesis. 
Similarly to the vector stochastic process defined in (\ref{eq:delvecstoc}),  
let
\beq
\bm{\Delta}^{(m)}(t)=\left(\Delta_1^{(m)}(t),..., \Delta_K^{(m)}(t)\right)
\label{eq:deltam}
\eeq
be the vector stochastic process containing the mean CSD introduced by tier $m$ for each tenant. Remarkably, $\Delta_i^{(m)}(t)$ is the stochastic process describing the $M/G/G_i^{(m)}(t)$ queue, that can be computed like in Section \ref{sect:queuemodel}, by replacing $\gamma \eta_i$ with $G_i^{(m)}(t) =  \sum_{\ell=1}^{L_m} G^{(m,\ell)}_{i}(t)$ in equations from (\ref{eq:sess}) to (\ref{eq:kingman}). 



Since the call flow traverses the service chain, the overall mean CSD is the sum of mean CSDs introduced by each single tier, namely
\beq
\boldsymbol{\Delta^c}(t) = \left( \Delta_1^{c}(t),..., \Delta_K^{c}(t) \right) =\sum_{m}\bm{\Delta}^{(m)}(t).
\label{eq:deltac}
\eeq

%
%
%
%
%

Similarly to the derivation of $\left\{p_{\boldsymbol{\eta}}, \bm{\delta}_{\boldsymbol{\eta}} \right\}$ previously obtained for a single CNF, $\bm{\Delta}^{(m)}(t)$ and $\bm{\Delta}^{c}(t)$ are ergodic CTMCs, and $\bm{\Delta}^{(m)}$  and $\bm{\Delta}^{c} $ correspond to their asymptotic behaviors as $t \rightarrow \infty$.
More technical details about the derivation of $\bm{\Delta}^{(m)}(t)$ and $\bm{\Delta}^{c}(t)$ (obtained by introducing the series and parallel structure functions) are provided in the Appendix A.

Accordingly, given $\bm{\delta}_{\boldsymbol{\eta}}^{(m)} = \left(\delta_{1, \boldsymbol{\eta} }^{(m)},...,\delta_{K, \boldsymbol{\eta} }^{(m)} \right)$ the mean delays vector  of tier $m$ in state $\boldsymbol{\eta}$, and $p_{\boldsymbol{\eta}}^{(m)} = \lim_{t\rightarrow\infty} \mathrm{Pr}\{\bm{\Delta}^{(m)}(t) = \bm{\delta}_{\boldsymbol{\eta}}^{(m)}\}$ the corresponding limiting probability, the set of pairs $\left\{p_{\boldsymbol{\eta}}^{(m)}, \bm{\delta}_{\boldsymbol{\eta}}^{(m)} \right\}$ represents the steady-state mean CSD distribution of tier $m$. Likewise, given $\bm{\delta}_{\boldsymbol{\eta}}^{c} = \left(\delta_{1,\boldsymbol{\eta}}^{c},...,\delta_{K,\boldsymbol{\eta}}^{c} \right)$ the mean delays vector of system in state $\boldsymbol{\eta}$, and $p_{\boldsymbol{\eta}}^{c} = \lim_{t\rightarrow\infty} \mathrm{Pr}\{\bm{\Delta}^{c}(t) = \bm{\delta}_{\boldsymbol{\eta}}^{c}\}$ the corresponding limiting probability, the set of pairs $\left\{p_{\boldsymbol{\eta}}^{c}, \bm{\delta}_{\boldsymbol{\eta}}^{c} \right\}$ is the steady-state mean CSD distribution of the entire service chain.

Letting $J^{(m)} = \prod_{\ell=1}^{L_m}N^{(m,\ell)}$ be the number of states of tier $m$ for each parallel CNF $\ell$, with $N^{(m,\ell)}$ given by (\ref{eq:numstati}), the number of states corresponding to the service chain is
\beq
J^c = \prod_{m \in \left\{ P, S, I, H \right\} }J^{(m)}.
\label{eq:VIMSstates}
\eeq
	
We consider the multi-tenant infrastructure as \textit{available} when {\em every} operator (or tenant) guarantees a mean CSD less than a (maximum) tolerated value for its customers. 

Let $\bm{W}^c(t)=\left( W_1^c(t),...,W_K^c(t) \right)$ be a $K$-dimensional vector containing the maximum tolerated values per tenant $i$ at time $t$. 
The {\em instantaneous availability} $A^{c}\left[t,\bm{W}^c(t)\right]$ is defined (see \cite{Levitin2003,Levibook}) as the probability that mean CSD of the service chain for each tenant $i$ (at $t>0$) is not greater than $W_i^c(t)$, $i=1,...,K$, viz.
\begin{equation}
A^{c}\left[t,\bm{W}^c(t)\right]=\mathrm{Pr}\{\Delta^c_i(t) - W_i^c(t)\leq 0, \;\; \forall i=1,...,K \}. 
\label{eq:at} 
\end{equation}
Given constant maximum values $\bm{W}^c(t) = \bm{w}^c=\left(w_1^c,...,w_K^c\right)$, the {\em steady-state availability} $A^{c}\left(\bm{w}^c\right)$ can be derived from (\ref{eq:at}) for $t\rightarrow\infty$, as  
\begin{eqnarray}
A^{c}(\bm{w}^c) {}={} \displaystyle\sum_{\boldsymbol{\eta} \in \Omega^c} p^{c}_{\boldsymbol{\eta}}\cdot
\bm{1}\left(\delta^{c}_{i,\boldsymbol{\eta}}\leq w_i^c, \forall i = 1,...,K\right),
\label{eq:astat2}
\end{eqnarray}
where $\bm{1}(\cdot)$ amounts to $1$ if condition holds true and $0$ otherwise, and $\Omega^c = \prod_m \Omega^{L_m}$.

\subsection{Steady-state availability using the MUGF technique}


The Multidimensional Universal Generating Function (MUGF) technique \cite{MDMTSC} provides an efficient method to evaluate the steady-state availability of (\ref{eq:astat2}). 
Being the MUGF a special case of probability generating function of a multivariate random variable, the steady-state distribution of an MSS can be expressed through a polynomial-shape form. More precisely, the MUGF of the steady-state mean CSD distribution pertaining to the tier $m$ is

\setlength{\arraycolsep}{0.0em}
\begin{equation}\label{eq:mugfnode}
u^{(m)}(\bm{z}) = \sum_{\boldsymbol{\eta} \in \Omega^{L_m}} p_{\boldsymbol{\eta} }^{(m)} \prod_{i=1}^{K}z_i^{\delta_{i,\boldsymbol{\eta} }^{(m)}},
\end{equation}
a function of the vector indeterminate $\bm{z}=(z_1,\dots, z_K)$.


From the generating functions theory, the sum of multivariate independent random variables has a generating function given by the product of the generating functions of single variables. 
Accordingly, by recalling that mean CSD of the service chain is the sum of mean CSDs introduced by each tier, the MUGF $u^{c}(\bm{z})$ is the product of the MUGFs of single tiers computed by (\ref{eq:mugfnode}), viz.
\beq
u^{c}(\bm{z}) =  \prod_{m} \left[    \sum_{\boldsymbol{\eta} \in  \Omega^{L_m}} p_{\boldsymbol{\eta}}^{(m)} \prod_{i=1}^{K}z_i^{\delta_{i,\boldsymbol{\eta}}^{(m)}}   \right].
\label{eq:MSFCsystem}
\eeq
Thus, $u^{c}(\bm{z})$ represents a polynomial-shape function in $z_1,\dots, z_K$. 
The $u^{c}(\bm{z})$ can be easily computed by combining the individual MUGFs through products and sums. 
The resulting expression of $u^{c}(\bm{z})$ provides the mean CSD vector $\bm{\delta}^{c}_{\boldsymbol{\eta}}$ and the corresponding steady-state probabilities $p^{c}_{\boldsymbol{\eta}}$ of the whole service chain. 
For each state $\bm{\eta}$, vector $\bm{\delta}^{c}_{\boldsymbol{\eta}}$ collects all the exponents of $z_1,\dots, z_K$, while $p^{c}_{\boldsymbol{\eta}}$ is the multiplicative coefficient.
Such quantities are used in (\ref{eq:astat2}) to compute the steady-state availability $A^{c}(\bm{w^c})$ of the multi-tenant chain.

\subsection{Redundancy optimization of the service chain}
\label{subsec:redundancy_optimization}

The proposed availability assessment method is useful to solve network design problems, such as the selection of an optimal configuration that satisfy a given availability objective. The problem of practical interest is to identify the configuration(s) minimizing the number of CNF replicas for each tier of the service chain.

We denote with the vector $\bm{\ell}$ a configuration of the multi-tenant service chain, that is, the number $L_{m}$ of replicas for each tier $m$. In the case of the cIMS, $m\in$ \{P, S, I, H\}, and $\bm{\ell}=\left(L_{P},L_{S},L_{I},L_{H}\right)$. Obviously, this approach can be easily applied to other SFC architectures by changing the set of elements belonging to the chain itself (namely the components in the series availability model).
 
We define $C^{(m,\ell)}$ as the cost of parallel CNF $\ell$ belonging to tier $m$. Thus, the cost of the configuration $\bm{\ell}$ of the multi-tenant service chain is
\beq
C^c(\bm{\ell})=\sum_{m} \sum_{\ell = 1}^{L_m} C^{(m,\ell)} .
\label{eq:cost}
\eeq

By considering $d_{max}$ the maximum tolerated value for the mean CSD for each tenant (which is typically provided by international standards, such as for the IMS \cite{itu1028}), namely $\bm{w}^c=\left(d_{max},\dots,d_{max}\right)$, the steady-state availability of the configuration $\bm{\ell}$, in terms of mean CSD, is given by (\ref{eq:astat2}), viz. 

\begin{equation}
A^{c}(\bm{w}^c,\bm{\ell}) = \displaystyle\sum_{\boldsymbol{\eta} \in J^c} p^{c}_{\boldsymbol{\eta}}\cdot
\bm{1}\left(\delta^c_{i,\boldsymbol{\eta}}\leq d_{max}, \forall i = 1,...,K\right) .
\label{eq:avdelay}
\end{equation}

Given an availability constraint $A_0$ (for example, $A_0=0.99999$ also known as ``five nines" availability), we define the set of configurations satisfying such a constraint as $\mathcal{L}^{c} = \left\{ \bm{\ell}:  A^{c}(\bm{w}^c,\bm{\ell})\geq A_0 \right\}$.  

Then, the system configurations with minimum deployment cost which satisfy the availability constraint $A_0$ are provided by solving the following optimization problem:    
\beq
\bm{\ell}^*= \argmin_{\bm{\ell} \in \mathcal{L}^{c} } C^c(\bm{\ell}) .
\label{eq:optprob}
\eeq
  
In summary, the optimization problem (\ref{eq:optprob}) contains elements both from availability and queueing models, where the latter allows to estimate the latency introduced by a chain. 
Indeed, the first step consists in computing the steady-state availability of chains, defined in terms of latency (see (\ref{eq:avdelay})), and in building the set of configurations $\mathcal{L}^{c}$ guaranteeing the given availability constraint. The second step consists in identifying the configuration(s) minimizing the cost (\ref{eq:cost}) represented by the solution(s) of (\ref{eq:optprob}). 

The process allowing to build various configurations relies on a greedy stage. By starting from a baseline configuration with $1$ CNF per tier, the routine automatically adds $1$ CNF per node up to a given threshold and evaluates the configuration availability. Since it is impossible to know beforehand which is the number of CNFs needed to obtain at least one solution, we set a configurable threshold value (namely, an integer number of CNFs) which represents the maximum number of CNFs a node can host. Obviously, depending on failure/repair parameters, it may happen that a given availability target (i.e., five nines) is never reached. In such a case, the network designer must relax the availability constraint (i.e. four nines). Once the routine has terminated its run, it produces a set of possible configurations, and it will choose the one ($\bm{\ell}^*$) satisfying the required condition. Interestingly, our routine allows to retain also the (sub-optimal) configurations, to leave the network designer the possibility of choosing a different setting (e.g. a configuration with $A^{c}(\bm{w}^c)=0.99998$, which is barely far from the five nines requirement).

\section{Experimental Results}
\label{sec:exp}

This section consists of two parts: the first one contains a detailed description of the testbed deployed to derive realistic parameters such as: repair rates of various components by adopting fault injection techniques, mean service times employed by containerized software instance to manage cIMS requests, mean call setup delays experimented by multimedia calls. The second part pertains to the availability assessment performed through MUGF technique by exploiting the estimated parameters.

\subsection{The cIMS testbed}

\begin{figure}[!t]
    \centering
    \includegraphics[width=0.62\columnwidth]{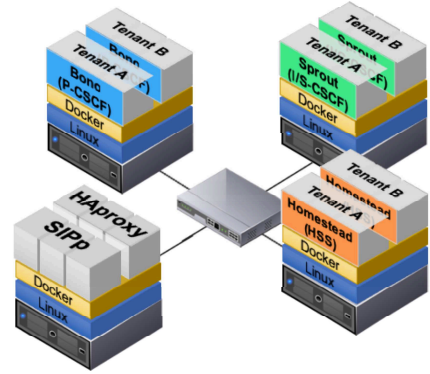}
    \caption{The deployed cIMS architecture.}
    \label{fig:cIMS_testbed_architecture}
\end{figure}

We deploy from scratch an experimental testbed aimed at validating the proposed technique in a realistic NFV-based environment. Our testbed is composed of hardware and software technologies commonly adopted in cloud datacenters such as: operating systems based on Linux kernel 4.4.0, Docker engines (version 19.03.5) running on 16-Core 1.80GHz Intel Xeon machines with 64GB RAM, two 500GB SATA HDD, four 1-Gbps Intel Ethernet NICs, and one NetApp Network Storage Array with 32TB of storage space and 4GB of SSD cache. The hosts are connected to a 1-Gbps Ethernet network switch. The testbed includes 4 machines (each of which equipped with the aforementioned HW/SW), and relies on Clearwater (release 130), an open-source platform (later embodied in a commercial product \cite{clearwater}) which allows emulating a fully working IMS architecture in a container-based environment. We deploy the IMS functionalities (Bono, Sprout, Homestead) on top of Docker as depicted in \figurename{}~\ref{fig:cIMS_testbed_architecture}. 
For each CNF we manage two different tenants. We use an external machine equipped with \textit{SIPp} tool \cite{sipptool} (as a workload generator) and with HAproxy \cite{haproxy} to perform traffic balancing among all instances of Bono. 

\begin{figure*}[!t]
	
	\centering
	\includegraphics[keepaspectratio=true, angle=270, width=0.5\textwidth]{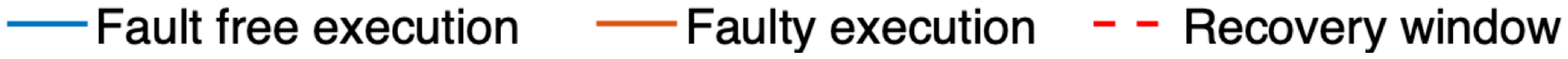}
	\subfloat[Software layer fault.\label{fig:sprout_ICSCF_injection_container}]{%
		\includegraphics[width=0.64\columnwidth]{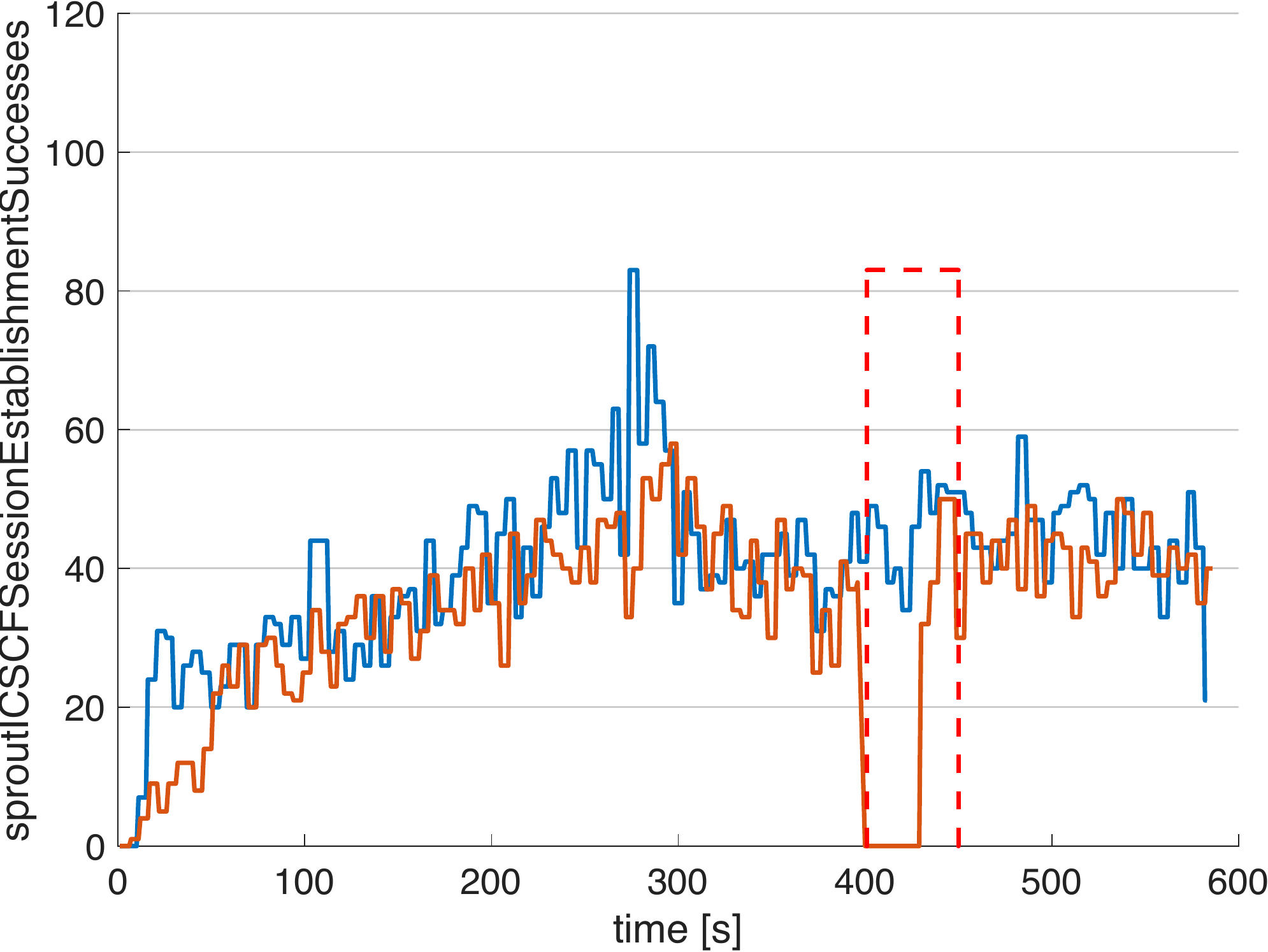}
	}
	\subfloat[Docker layer fault.\label{fig:sprout_ICSCF_injection_dockerd}]{%
		\includegraphics[width=0.64\columnwidth]{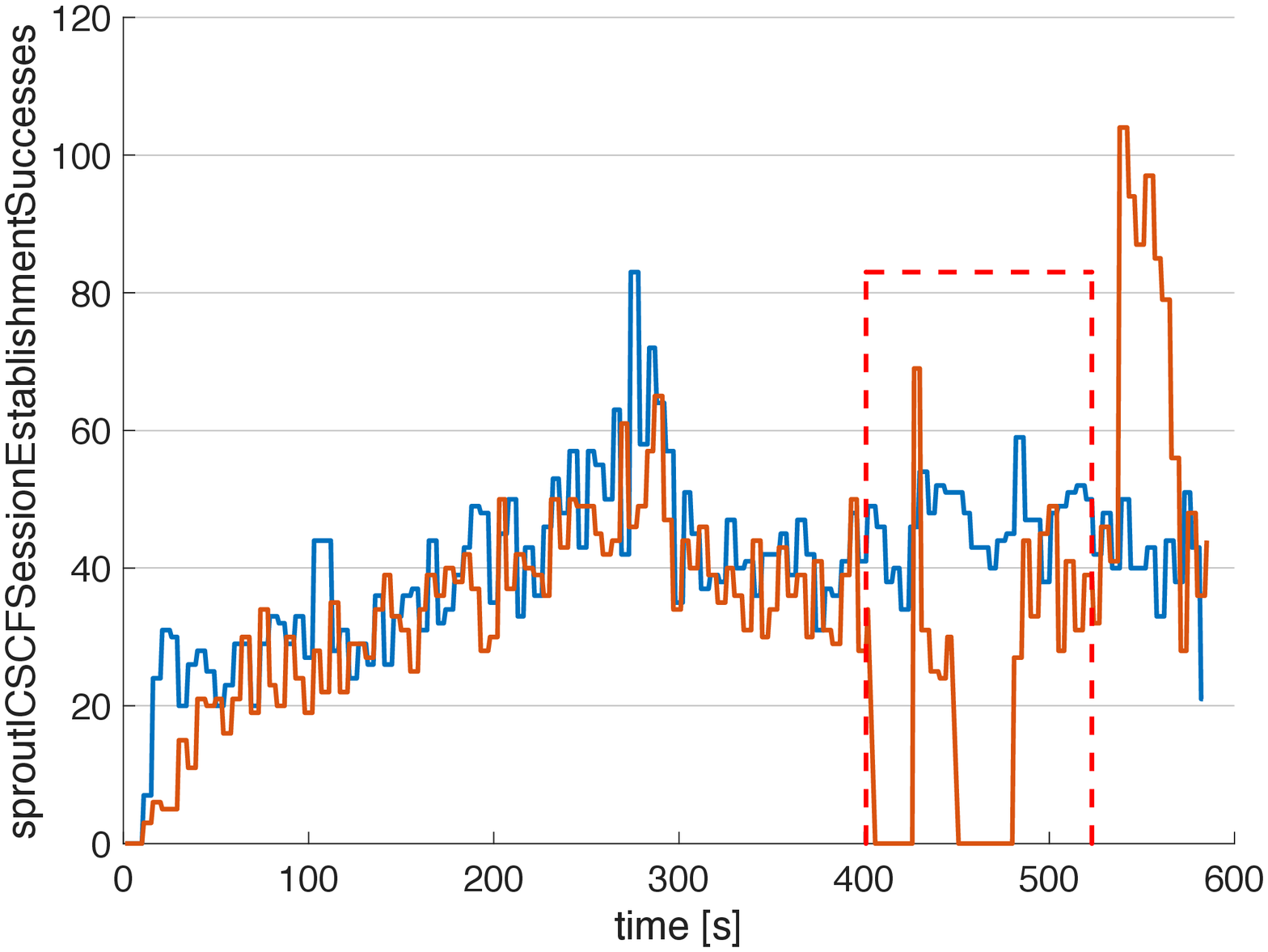}
	}
	\subfloat[Infrastructure layer fault.\label{fig:sprout_ICSCF_injection_infrastructure}]{%
		\includegraphics[width=0.65\columnwidth]{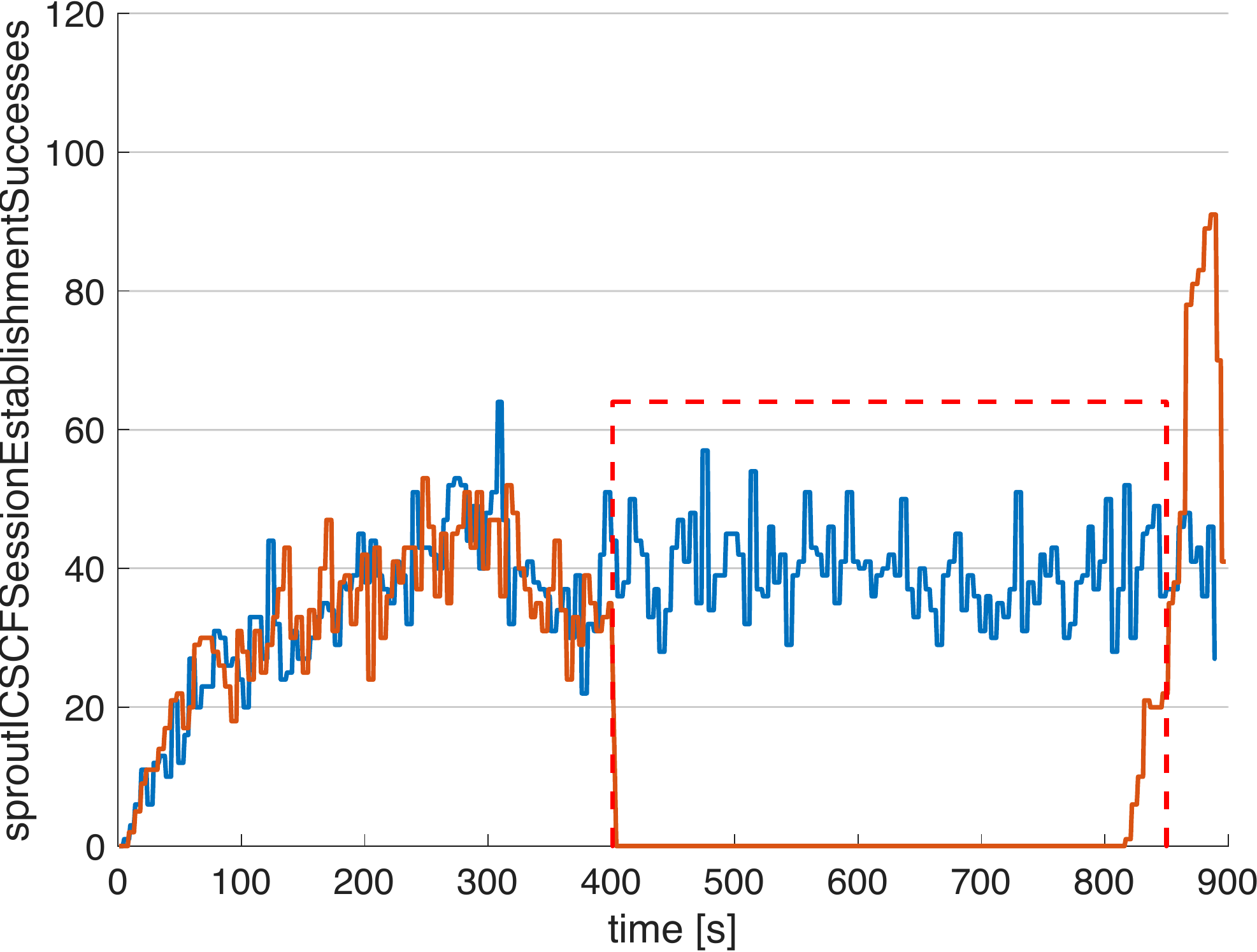}
	}
	
	\caption{Sprout I-CSCF under fault injection.}
	\label{fig:sprout_icscf_full}
\end{figure*}

\begin{figure*}[t!] 
	\centering
	\begin{tabular}{cccc}
		\subfloat{\includegraphics[scale=0.22]{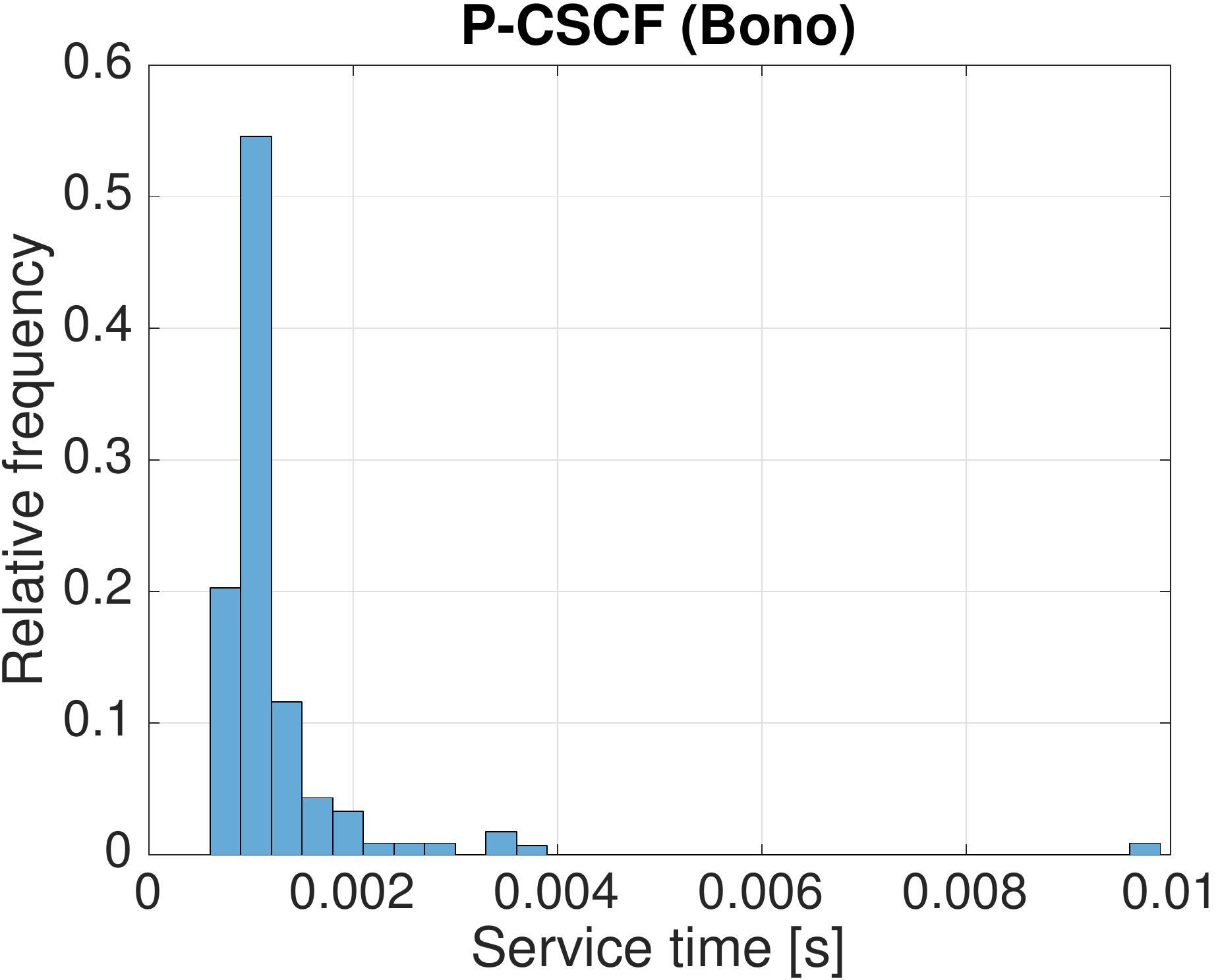}}  \hspace{2mm}
		\subfloat{\includegraphics[scale=0.22]{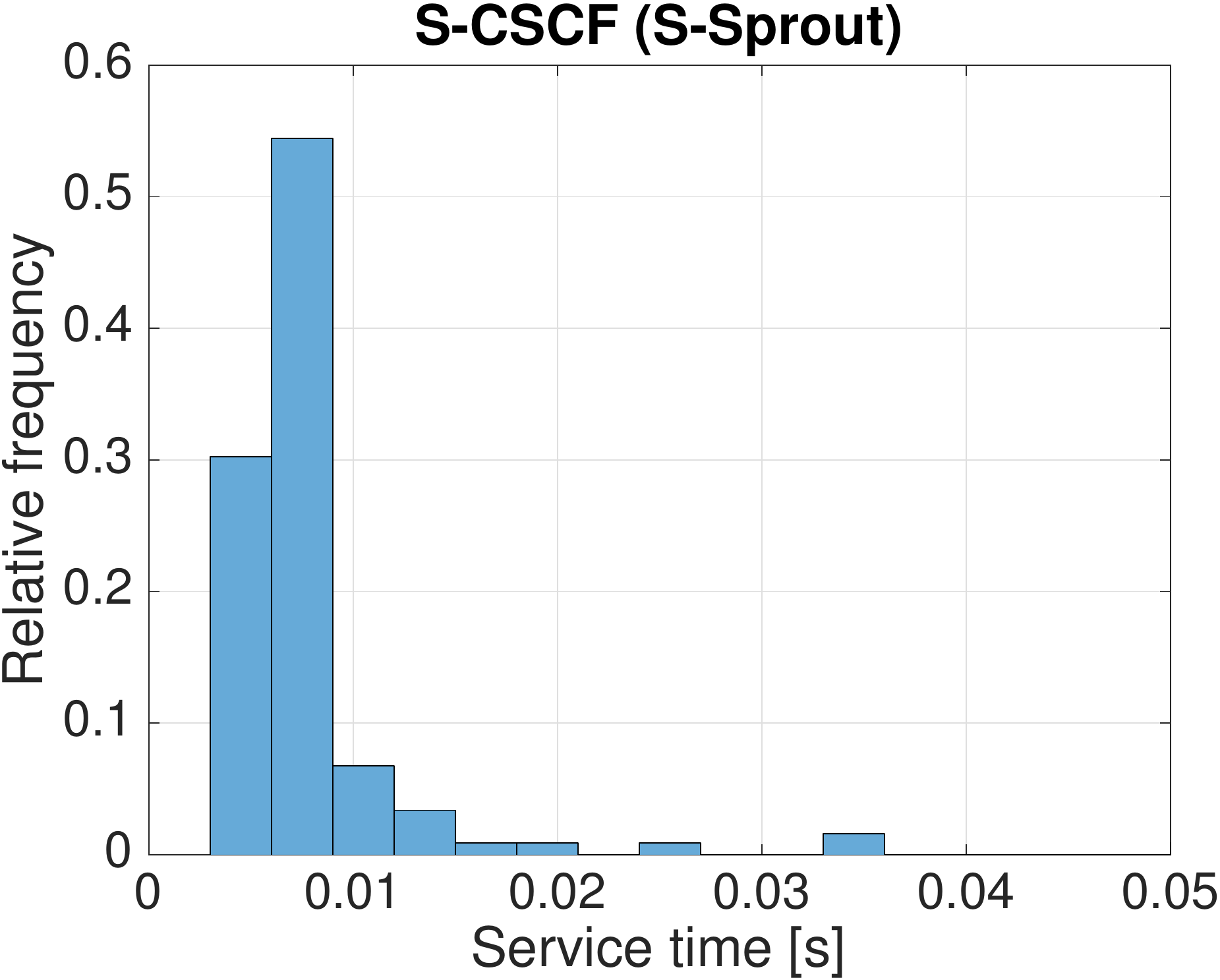}} \hspace{2mm}
		\subfloat{\includegraphics[scale=0.22]{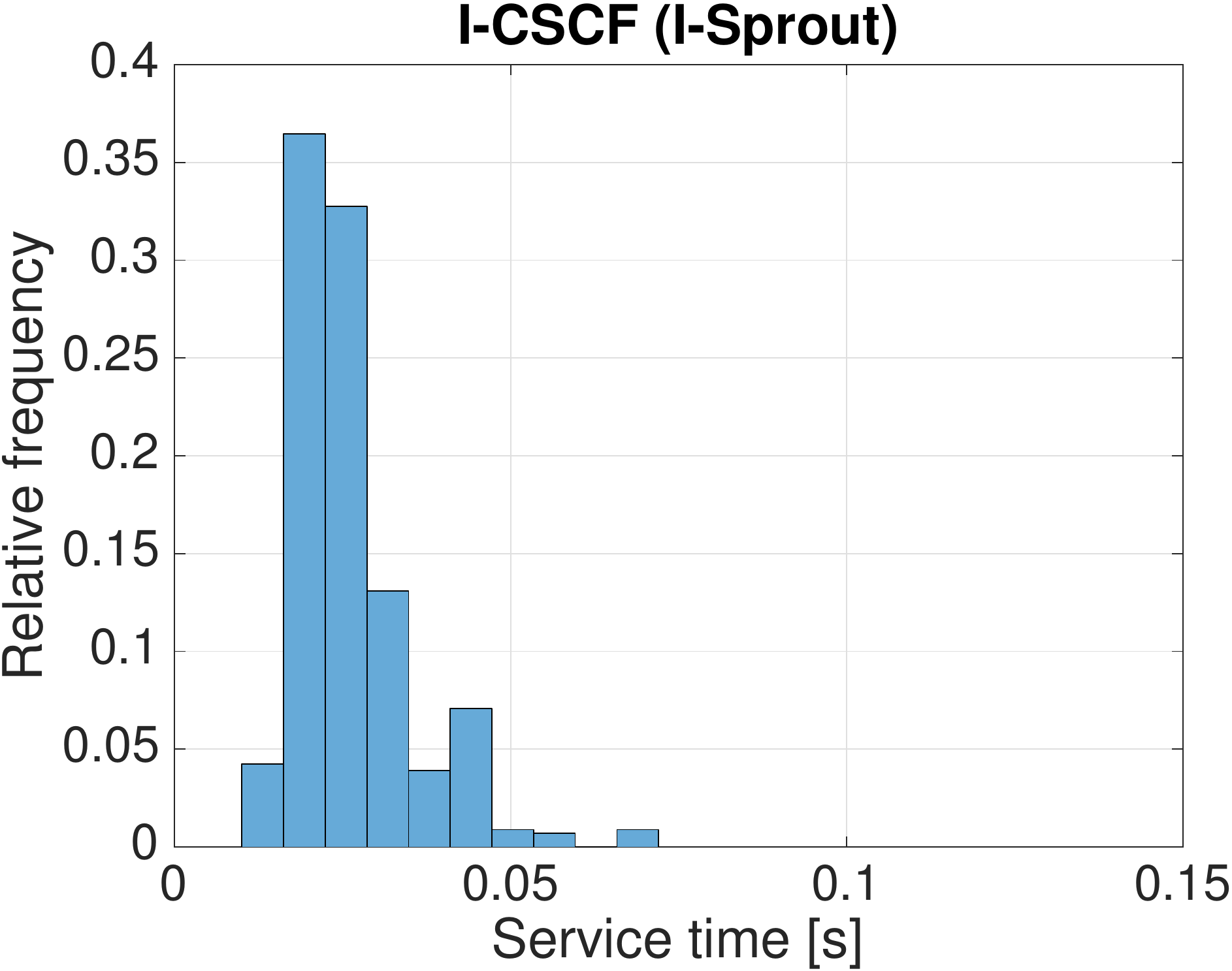}} \hspace{2mm}
		\subfloat{\includegraphics[scale=0.22]{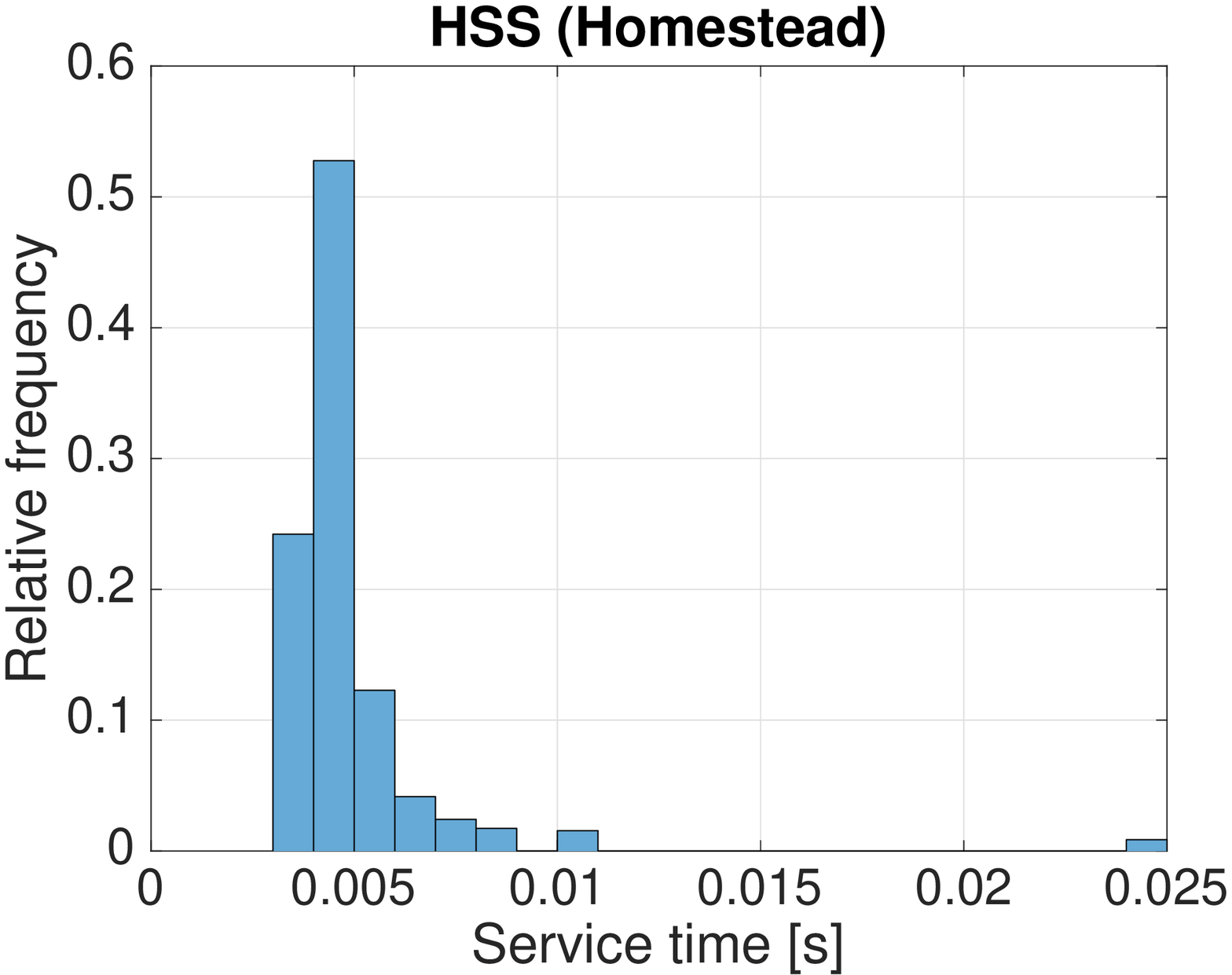}} 
	\end{tabular}
	\caption{Empirical service time distribution of Clearwater nodes (from the left: P-CSCF, S-CSCF, I-CSCF, HSS).}
	\label{fig:pdfservtimes}
\end{figure*}

As in our previous studies \cite{cotroneo17,androfit}, we adopt \emph{fault injection} to emulate faults and to measure the recovery times, in order to estimate representative model parameters. 
To assure the occurrence of failures, we emulate faults through the injection of their effects, which is also referred to as \emph{error} or \emph{failure injection} in some studies \cite{avizienis2004basic,natella2016assessing}. In our experiments, we injected the following three types of faults.

\textbf{Software layer faults}: responsible for software crashes of the CNF upper layer which embeds the specific IMS service logic. Typically, such faults include race condition bugs, resource exhaustion due to software aging bugs, I/O and exception handling bugs \cite{grottke2007fighting}. The Clearwater IMS is no exception, as a number of such failures have been reported by users on its issue tracker and mailing lists \cite{clearwater2016mailing}. 
We inject these faults by forcing the abrupt termination of a container.

\textbf{Docker layer faults}: similarly for the containers, the Docker engine is affected by software faults related to timing, resource management, and other environmental conditions. Both academic research and end-users report recurring failures in Docker and in similar management software \cite{torquato2019experimental,machida2012aging,matos2012experimental}, causing the unavailability of containers along with virtual networks and storage volumes. We inject these faults by forcing the termination of container management services (i.e., the {\lmttfont dockerd} process) that, in turn, results in the termination of all containers running on top.

\textbf{Infrastructure layer faults}: software and hardware faults related to the crash of the hypervisor and the underlying physical infrastructure, respectively \cite{machida2012aging,cotroneo17}. In turn, these faults cause the unavailability of Docker and the whole set of containers. We inject these faults by forcing the abrupt shutdown of the machine.

We performed 30 fault injection experiments per CNF and per fault type, amounting to 360 experiments in total. Each fault injection experiment takes about 10 minutes both for the software layer and for the Docker engine, whereas it takes about 15 minutes for the infrastructure layer. Before injecting faults, we wait for an initial warm-up period (400 seconds) to let the system to reach a regime level. We automate fault injection experiments through ad-hoc routines developed to manage operations such as: start/stop fault injection, trigger the shutdown and the recovery of containers and hosts, collect metrics for each CNF through SNMP protocol as detailed in the following.

\begin{itemize}

    \item \textbf{P-CSCF}: we analyze the number of SIP events (SIP messages) successfully passed to a Bono worker thread per unit time, reported in the SNMP object {\lmttfont bonoQueueSuccessFailSuccesses}.

    \item \textbf{S-CSCF}: we analyze the number of successful outgoing SIP transactions (INVITE messages) per unit time, reported in the SNMP object {\lmttfont sproutSCSCFOutgoingSIPTransactionsSuccesses}.

    \item \textbf{I-CSCF}: we analyze the number of successful terminating request attempts over the period, reported in the SNMP object {\lmttfont sproutICSCFSessionEstablishmentSuccesses}.

    \item \textbf{HSS}: we analyze the number of enqueued Memcached requests per unit time, reported in the SNMP object {\lmttfont homesteadCacheQueueSizeCount}.

\end{itemize}

We use the aforementioned metrics to estimate the \emph{recovery time} of the cIMS, in line with the previous NFV dependability benchmark study \cite{cotroneo17}. We measure the recovery time as the period during which a CNF is unavailable: it starts at a given fault injection time when the considered metric drops to zero, and it ends when the metric raises up to its regime value after recovery is completed. 
Each metric is evaluated at the output of a five-sample moving average filter, introduced to smooth fluctuations occurring during recovery. We consider the recovery procedure as completed when the metric overcomes a given threshold set to $90\%$ of the fault-free metric level \cite{kanoun2008dependability}.

For instance, the recovery times of Sprout I-CSCF CNF after different injections are shown in panel of Figs. 8. Precisely, \figurename{}~\ref{fig:sprout_ICSCF_injection_container}, \figurename{}~\ref{fig:sprout_ICSCF_injection_dockerd}, \figurename{}~\ref{fig:sprout_ICSCF_injection_infrastructure} report the recovery time in the presence of a Software layer fault, a Docker layer fault, and an infrastructure layer fault, respectively. We note that the differences between the curves is due to the typical variability of the experiments. Moreover, it is possible to receive a slightly different amount of traffic across executions, since load balancing cannot be perfectly uniform across replicas.
The panel of Figs. \ref{fig:pdfservtimes} reports the service times distributions that we measure for each cIMS node. 
We remarked that these empirical distributions of service times are important to analyze the latency, namely to derive the coefficients of variation for each CNF to be used in (\ref{eq:kingman}).

Interestingly, our fault injection trials provided unexpected findings on the failure and recovery behavior of containerized network functions. Precisely, most studies adopt too simplistic assumptions about the containers time-to-recovery, by only considering the time to perform a restart action on a container (in the order of few seconds for a Linux container on a modern hardware machine). In contrast, our experiments reveal a longer time (in the order of minutes) to restore performance. A critical factor is represented by the restart of application software. Since the IMS is developed in the Java language, a new instance of the JVM needs to be allocated and initialized. Moreover, the application itself needs to manage allocations and data initialization (e.g., to start a thread pool). Afterwards, the performance is gradually restored, due to enqueueing and caching effects.

\subsection{Availability assessment of cIMS}

We perform a steady-state availability assessment of cIMS through the MUGF technique supported by the experimental data. 
Let us assume that: \textit{i)} each CNF composing the system in Fig. \ref{fig:serpar} exhibits one and the same availability model namely, we suppose identical failure/repair parameters for each CNF), and, \textit{ii)} all nodes exhibit one and the same cost amounting to $1$, or equivalently, $C^{(m,\ell)} = 1$, $\forall \ell \in \left\{1,...,L_m\right\}$, with $m \in$ $\left\{ \textnormal{P},\textnormal{S},\textnormal{I},\textnormal{H} \right\}$. Needless to say, the considered assumptions can be easily tailored across a variety of scenarios where network providers operate with different costs and performance features. 
Let us refer to an exemplary setting with $K=2$ network providers (tenants), where the first one has $n_1=2$ containerized instances, and the second one has $n_2=3$ containerized instances.  
As a result, the MSS model of the CNF can be directly derived from the transition-state diagram depicted in Fig. \ref{fig:miniseph} with a number of different states amounting to $N = (n_1+1)(n_2+1) + 2 = 14$ according to (\ref{eq:numstati}). The steady-state probability distribution of a single CNF can be directly obtained by solving the system of differential equations (\ref{eq:diff_eq}) for $t\rightarrow\infty$, as detailed in Appendix B.

Table \ref{tab:params} summarizes the input parameters adopted for the assessment, where: failure and repair rates have been, in part, derived from the deployed testbed, and, in part, chosen according to the technical literature (see \cite{matos2012,sebastio18,hennpatt,kleppmann,usenix,trivediaccess}). Let us provide clarifications about some parameters estimated through the testbed. First, the presence of multiple containers does not dramatically affect the infrastructure reboot ruled by the $\mu_I$ parameter. Such a situation is commonly encountered in practice, where the network designer adopts some strategies to control the parameter variability (i.e., smart allocation of hardware resources, snapshots usage, CPU pinning). 

Moreover, the $\alpha$ parameters are influenced by the capacity of the system. We calibrated the workload generator to run new subscribers and to generate traffic such that the system approaches its capacity, without degrading latency and without saturating CPU and memory. This scenario assumes that the number of containers in the system is scaled according to the workload, which is typical for services deployed on cloud computing infrastructures.
Again, the empirical service times distributions (Figs. \ref{fig:pdfservtimes}) are evaluated by analyzing the service rates of each node. 
From such data, we also derive the coefficient of variation per node, useful to obtain the Kingman's approximation (\ref{eq:kingman}).
Finally, the {\em d}$_{max}$ value is a pessimistic estimate chosen in accordance to the ITU-T standard specifications \cite{itu1028}, where acceptable values are in the order of seconds, since they account for propagation delays in geographic networks that are obviously negligible in our testbed.

\begin{table}[t!]
	\caption {Input parameters}
	\label{tab:params}
	\resizebox{.48\textwidth}{!}{
		\begin{threeparttable}
			\begin{tabular}{c|c|c}
				\hline
				Parameter & Description & Value\\
				\hline
				\\[-8pt]
				$1/\lambda_{C}$ & mean time for container failure\tnote{$\dagger$} & 1258 hours \\ \hline
				$1/\lambda_{D}$ & mean time for docker failure\tnote{$\dagger$} & 2516 hours \\ \hline
				$1/\lambda_{I}$ & mean time for infrastructure failure\tnote{$\dagger$} & 60000 hours \\ \hline
				$1/\mu_{C}$ & mean time for container repair\tnote{$\ddagger$} & 30 s\\ \hline
				$1/\mu_{D}$ & mean time for docker repair\tnote{$\ddagger$} & 60 s \\ \hline
				$1/\mu_{I}$ & mean time for infrastructure repair\tnote{$\ddagger$}  & 5 min \\ \hline
				$\alpha_1$ & IMS request arrival rate at tenant $1$\tnote{$\ddagger$}& 100 s$^{-1}$ \\ \hline
				$\alpha_2$ & IMS request arrival rate at tenant $2$\tnote{$\ddagger$}& 200 s$^{-1}$ \\ \hline
				$1/\beta_P$ & P-CSCF empirical mean service time per request\tnote{$\ddagger$} & $1.1 \cdot 10^{-3}$ s \\ \hline
				$1/\beta_S$ & S-CSCF empirical mean service time per request\tnote{$\ddagger$} & $7.2 \cdot 10^{-3}$ s \\ \hline
				$1/\beta_I$ & I-CSCF empirical mean service time per request\tnote{$\ddagger$} & $4.1 \cdot 10^{-2}$ s \\ \hline
				$1/\beta_H$ & HSS empirical mean service time per request\tnote{$\ddagger$} & $4.6 \cdot 10^{-3}$ s \\ \hline
				$CV_P$ & P-CSCF coefficient of variation (Kingman's approx.)\tnote{$\ddagger$} & $0.7538$ \\ \hline
				$CV_S$ & S-CSCF coefficient of variation (Kingman's approx.)\tnote{$\ddagger$} & $0.9826$ \\ \hline
				$CV_I$ & I-CSCF coefficient of variation (Kingman's approx.)\tnote{$\ddagger$} & $0.5581$ \\ \hline
				$CV_H$ & HSS coefficient of variation (Kingman's approx.)\tnote{$\ddagger$} & $0.4631$ \\ \hline
				{\em d}$_{max}$ & Maximum tolerated CSD & $50$ ms \\ \hline
				\hline
			\end{tabular}
			\begin{tablenotes}
				\item [$\dagger$] From scientific literature
				\item [$\ddagger$] From experiments
			\end{tablenotes}
		\end{threeparttable}
	}
\end{table} 

	\begin{table}[t!] 
		\caption {Steady-state Availability under $12$ configurations} \label{tab:exemplarysettings}
		\centering
		\resizebox{.48\textwidth}{!}{
			\begin{tabular}{c|c|c|c}
				\hline
				Config. & Redundancy Level & $C^c(\bm{\ell})$ & $A^{c}(\bm{w}^c)$ \\
				\hline
				$\bm{\ell^*}$ & $[CNF^{(P)}=2, CNF^{(S)}=1, CNF^{(I)}=3 ,CNF^{(H)}=2]$ & 8 & 0.999992  \\ \hline
				$\bm{\ell_1}$ & $[CNF^{(P)}=2, CNF^{(S)}=2, CNF^{(I)}=2 ,CNF^{(H)}=2]$ & 8 & 0.999944  \\ \hline
				$\bm{\ell_2}$ & $[CNF^{(P)}=2, CNF^{(S)}=3, CNF^{(I)}=2 ,CNF^{(H)}=2]$ & 9 & 0.999944  \\ \hline
				$\bm{\ell_3}$ & $[CNF^{(P)}=3, CNF^{(S)}=3, CNF^{(I)}=2 ,CNF^{(H)}=3]$ & 11 & 0.999945  \\ \hline
				$\bm{\ell_4}$ & $[CNF^{(P)}=1, CNF^{(S)}=1, CNF^{(I)}=3 ,CNF^{(H)}=3]$ & 8 & 0.999984   \\ \hline
			    $\bm{\ell_5}$ & $[CNF^{(P)}=1, CNF^{(S)}=1, CNF^{(I)}=2 ,CNF^{(H)}=1]$ & 5 & 0.999919\\ \hline
			    $\bm{\ell_6}$ & $[CNF^{(P)}=2, CNF^{(S)}=1, CNF^{(I)}=2 ,CNF^{(H)}=1]$ & 6 & 0.999927\\ \hline
	            $\bm{\ell_7}$ & $[CNF^{(P)}=2, CNF^{(S)}=2, CNF^{(I)}=2 ,CNF^{(H)}=1]$ & 7 & 0.999936\\ \hline
	            $\bm{\ell_8}$ & $[CNF^{(P)}=3, CNF^{(S)}=3, CNF^{(I)}=3 ,CNF^{(H)}=1]$ & 10 & 0.999994\\ \hline	    
				$\bm{\ell_9}$ & $[CNF^{(P)}=2, CNF^{(S)}=2, CNF^{(I)}=3 ,CNF^{(H)}=2]$ & 9 & 0.9999999  \\ \hline
			    $\bm{\ell_{10}}$ & $[CNF^{(P)}=2, CNF^{(S)}=2, CNF^{(I)}=2 ,CNF^{(H)}=4]$ & 10 & 0.999968   \\ \hline 
				$\bm{\ell_{11}}$ & $[CNF^{(P)}=2, CNF^{(S)}=3, CNF^{(I)}=2 ,CNF^{(H)}=4]$ & 11 & 0.999968  \\ \hline
				\hline
			\end{tabular}
		}
		\label{tab:avavalues}
	\end{table}

\begin{figure*}[!t]
	\setcounter{equation}{23}
	\setlength{\arraycolsep}{0.0em}
		\begin{scriptsize}
	\begin{eqnarray}
	u^{c}(\bm{z}) &{}={}& 0.9997 \; {z_{1}}^{0.0255} {z_{2}}^{0.0255} \; + 2.649 \times10^{-5} \; {z_{1}}^{0.0255} \; {z_{2}}^{0.0255}  \; + 8.773 \times10^{-11} \; {z_{1}}^{0.0255} \; {z_{2}}^{0.0255}  \nonumber\\
	&&{+}\: 1.135 \times10^{-29} \; {z_{1}}^{0.0304} {z_{2}}^{0.0329} +  4.013 \times10^{-34} \; {z_{1}}^{0.0417} {z_{2}}^{0.0329} \;\; + \dots + \dots \nonumber\\
	&&{+}\: \textcolor{red}{1.061 \times10^{-44} \; {z_{1}}^{0.0587} {z_{2}}^{0.0329}} +  2.121 \times10^{-44} \; {z_{1}}^{0.0255} {z_{2}}^{0.0329} \;\; + \dots + \dots \nonumber\\
	&&{+}\: \textcolor{red}{3.498 \times10^{-49} \; {z_{1}}^{0.0450} {z_{2}}^{0.0589}} +  \textcolor{red}{2.867 \times10^{-74} \; {z_{1}}^{0.0416} {z_{2}}^{0.0589}} \;\; + \dots + \dots  \nonumber\\
	&&{+}\: \textcolor{red}{2.142 \times10^{-58} \; {z_{1}}^{0.0418} {z_{2}}^{0.0570}} +  1.654 \times10^{-75} \; {z_{1}}^{0.0281} {z_{2}}^{0.0137} \;\; + \dots + \dots 
	\label{eq:supermugf}
	\end{eqnarray}
			\end{scriptsize}
	\setlength{\arraycolsep}{5pt}
	\hrulefill
\end{figure*}

\begin{figure}[t!]
	\centering
	\captionsetup{justification=centering}
	\includegraphics[scale=0.3,angle=90]{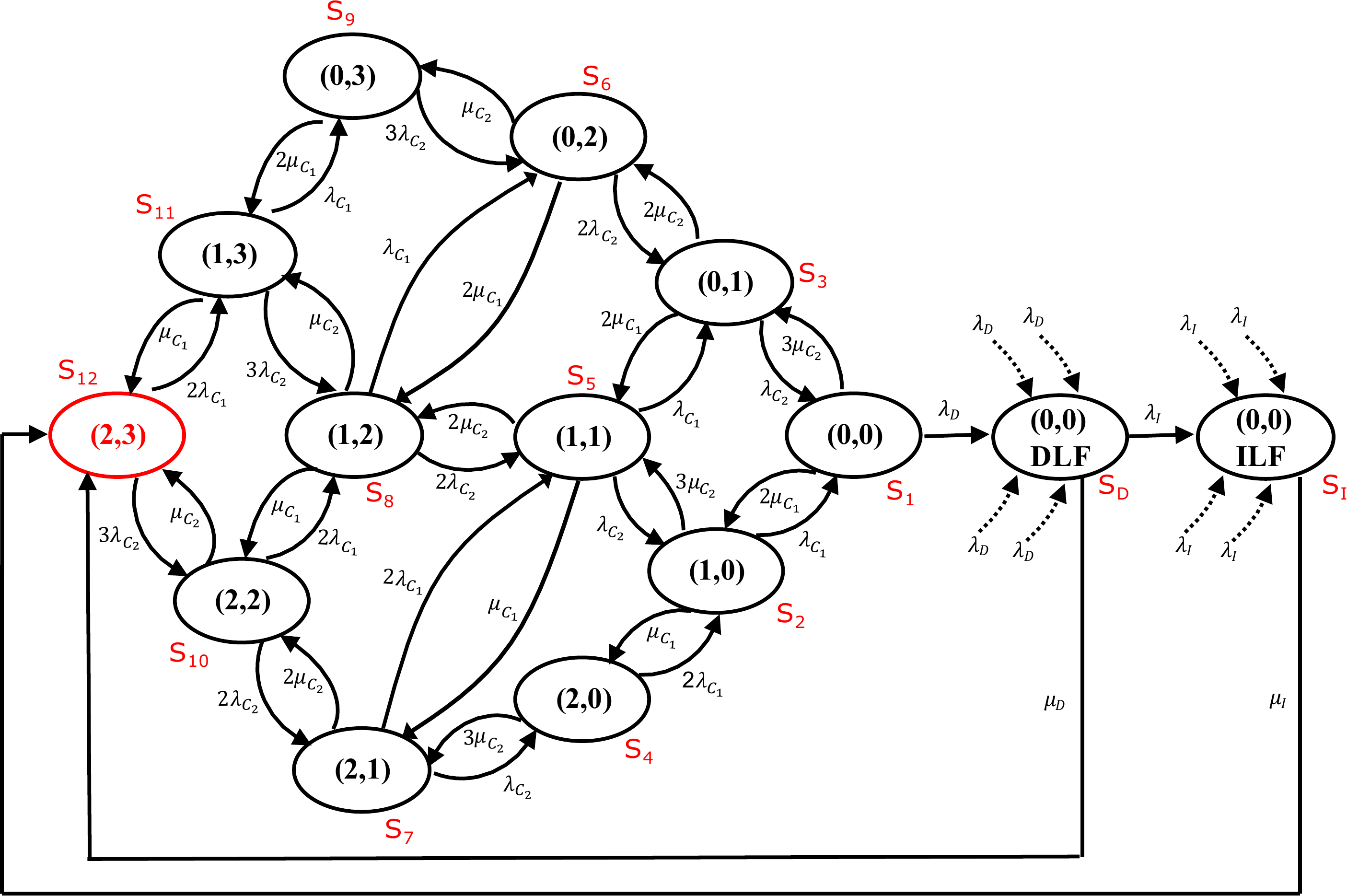}
	\caption{Transition-state diagram of the CNF with two tenants.}
	\label{fig:miniseph}
\end{figure} 

In keeping with the ``five nines" availability requirement, we set $A_0 = 1 - 10^{-5}$ and solve the optimization problem (\ref{eq:optprob}). A routine in Mathematica\textsuperscript{\textregistered} (available upon request) computes the MUGF (\ref{eq:MSFCsystem}) and  serves to finally evaluate the steady-state availability of cIMS expressed through (\ref{eq:astat2}). Then, the routine selects, among all feasible configurations, the one(s) exhibiting the minimum cost. 

Table \ref{tab:avavalues} reports the results of our experimental evaluation, where $12$ exemplary configurations have been shown. The optimal configuration is $\bm{\ell^*}$, which exhibits a steady-state availability amounting to $A^{c}(\bm{w}^c) = 0.999992$, with a cost amounting to $C^c(\bm{\ell^*})=8$ CNFs.
Such a configuration is obtained by considering (see the second column of Table \ref{tab:avavalues} where CNF$^{(m)}$ denotes the number of CNFs for tier $m$): $2$ redundant CNFs for P-CSCF and HSS, $3$ CNF replicas for I-CSCF, and no redundancy for the S-CSCF. For the sake of clarity, we want to highlight that the same cost and availability values are obtained by exchanging the redundancy role of P-CSCF, S-CSCF, and HSS. This is due to the fact that the most critical network function turns to be the I-CSCF, in view of its higher service time (see pertinent values in Table \ref{tab:params}).   
By exploring the remaining configurations, we can observe other interesting facts. 
Configuration $\bm{\ell_1}$ is obtained as a rearrangement of $\bm{\ell^*}$, where replicas have been differently distributed across the network functions (obviously, this results in the same cost). But, as can be noticed, this redistribution does not allow to meet the desired high availability requirement.  
In configuration $\bm{\ell_2}$, the redundancy of S-CSCF is empowered by $2$ replicas w.r.t. $\bm{\ell^*}$, whereas one less replica is considered for I-CSCF. Also in this case, the steady-state availability fixes on ``four nines", and the overall configuration cost increases to $9$. 
Interestingly, configurations $\bm{\ell_1}$ and $\bm{\ell_2}$ exhibit the same value of availability even if an additional S-CSCF replica characterizes $\bm{\ell_2}$ w.r.t. $\bm{\ell_1}$. This behavior can be ascribed to the high efficiency of S-CSCF in handling IMS requests (see $1/\beta_S$ value in Table \ref{tab:params}), translating in a very scarce sensitivity in improving its availability when the number of the corresponding CNF replicas exceeds the value of $2$.
	
Even adding one more replica to P-CSCF and to HSS w.r.t. $\bm{\ell_2}$, the steady-state high availability target is not reached, and the whole cost jumps to $11$ (configuration $\bm{\ell_3}$). Again, such a behavior can be ascribed to the weakness introduced by I-CSCF. This notwithstanding, in case we preserve a high redundancy degree for I-CSCF with $3$ replicas (as occurs for the optimal configuration $\bm{\ell^*}$), the availability target remains unsatisfactory if two nodes are not replicated at all as shown in $\bm{\ell_4}$. Thus, guaranteeing a strong redundancy degree for the I-CSCF it is not enough. 

A set of ``cheap'' configurations includes $\bm{\ell_5}$ - $\bm{\ell_7}$, whose costs range from $5$ to $7$. They show that when a more relaxed availability constraint (e.g. ``four nines") has to be satisfied, it might be no necessary to add many CNF replicas. A limiting case is given by $\bm{\ell_5}$, whose availability value amounts to $0.999919$ at a very cheap cost of $5$. In contrast,  the ``five nines" requirement is met by $\bm{\ell_8}$ which cannot be the optimal configuration since its cost amounts to $10$ (more expensive than $\bm{\ell^*}$).
Furthermore, configuration $\bm{\ell_9}$ is obtained by adding one more replica to S-CSCF w.r.t. $\bm{\ell^*}$. In such a case, cost increases by one, but, the steady-state availability jumps to 0.9999999, a very challenging value sometimes required by mission-critical services. Interestingly, $\bm{\ell_9}$ exhibits a greater availability value than $\bm{\ell_8}$, but at a cheaper cost (amounting to $9$). This behavior depends on the fact that in $\bm{\ell_8}$ no redundant CNF is allocated to HSS, and confirms that the best trade-off between availability and costs can be obtained through smart allocation of CNF replicas among the tiers.
Finally, configurations $\bm{\ell_{10}}$ and $\bm{\ell_{11}}$ report two cases of CNF redundancy greater than $3$ (for a tier). Precisely, $\bm{\ell_{10}}$ can be considered as an enhancement of configuration $\bm{\ell_{1}}$ with $2$ more replicas on HSS tier. Likewise, $\bm{\ell_{11}}$ is derived by $\bm{\ell_{2}}$ adding $2$ more CNFs to the HSS. In both cases, we do not observe any significant improvement in the availability values (both amounting to $0.999968$ for $\bm{\ell_{10}}$ and $\bm{\ell_{11}}$) which are still far from the five nines availability target. Even, the costs of $\bm{\ell_{10}}$ and $\bm{\ell_{11}}$ increase by two units with respect to $\bm{\ell_{1}}$ and $\bm{\ell_{2}}$, respectively. Once again, we can notice that a wrong redundancy strategy contributes to make the costs grow but not the availability values.

\noindent In (24) we report the MUGF of cIMS system in the optimal configuration $\bm{\ell^*}$, where most terms have been suppressed due to space constraints. The first term has a coefficient equal to $0.9997$, which is the steady-state probability for a state corresponding to a mean CSD equal to $0.0255$ s for both tenants, as indicated by the exponents of $z_1$ and $z_2$. Moreover, the terms in red refer to states where one or both mean CSD values do not respect the mean CSD constraint $d_{max}$, namely are greater than $50$ ms, and their coefficient does not contribute to the value of the steady-state availability of cIMS according to (\ref{eq:astat2}).


From a computational complexity perspective, it is useful to highlight that the MUGF approach mitigates by far the computational load required by monolithic approaches which attempt to find the steady-state probability distribution of a single CTMC describing the whole system without decomposing it in simpler subsystems. 
Indeed, as regards the proposed example  with $N^{(m,\ell)}=N=14$, the state space of a single cIMS CTMC model amounts to  $J^c = 14^{\sum_{m \in \{ {P,S,I,H\}}} L_m}$  (by virtue of (\ref{eq:VIMSstates})). In particular, the optimization problem (\ref{eq:optprob}) requires to solve $r^m$ systems of equations whose number ranges from $14^4$ to $J^c$, being $r$ the maximum redundancy level considered in the optimization algorithm. For instance, the optimal configuration  $\bm{\ell^*}=(2, 1, 3, 2)$ requires the solution of a system with $14^{ 2 + 1 + 3 + 2 } = 14^8$ equations.
Conversely, our approach requires three steps: $i)$ we find the steady-state distribution of the CTMC of a single CNF with $14$ states (namely we need to solve a system of $14$ equations for each tier as detailed in Appendix B); $ii)$ we compute the steady-state distribution of the mean CSD due to tier $m$, for each $m$, and the corresponding MUGF by (\ref{eq:mugfnode}); $iii)$ we combine the distributions computed in step $ii)$ to obtain the mean CSD probability distribution of cIMS. Then, we get the steady-state availability (\ref{eq:astat2}).
To compute the MUGF corresponding to the optimal configuration for the cIMS, we need about $360$ s on a PC equipped with an Intel Quad-Core Xeon E5 CPU@3.7GHz.


\subsection{Limitations}
\label{sec:limitation}
This work presented an availability model for multi-tenant service chains, tailored to the recent architectural trend towards containerized services. Although our efforts to match real systems (e.g., non-exponential service times assumptions, parameters from real-world experiments, etc.), some limitations necessarily remain: $i)$ for the sake of simplicity, in our model we consider CNFs having the same performance, but in real architectures, they might slightly differ. This notwithstanding, this assumption holds for many systems (including our cIMS case study), where all of the CNFs have been developed using the same software technology, have been assigned the same amount of resources (e.g., in terms of virtual CPUs), and share the same underlying layers (Docker and Infrastructure) in a cloud-based deployment; $ii)$ some parameters (e.g., the failure rates) are derived from the technical literature rather than from experiments, due to the large observation scale of failure events (up to some years). The other parameters are estimated using well-assessed fault injection techniques; $iii)$ we assume exponential distributions for failure/repair times: such an assumption is the most common and accepted across the technical literature, and in our work, it is necessary to manage complications arising from the joint availability and queueing modeling. 
Moreover, our proposed method sacrifices some high-level expressiveness achieved by other approaches (e.g., SPN/SRN), in order to benefit from visibility on the underlying analytical model. For example, this occurs in the eq. (18) where the MUGF expression of the mean delay distribution pertaining to the tier $m$ is made explicit in terms of the pair $(p_{\boldsymbol{\eta}}, \bm{\delta}_{\boldsymbol{\eta}})$. Benefiting from such a decomposition, the MUGF of the overall chain can be easily evaluated through a simple product as shown in eq. (19). 
Finally, we note that the MUGF method is intended to be applied ``one-shot'', using a fixed set of parameters, reflecting the expected workload, mean time to failure, mean time to repair, etc. 
This typically reflects the SLA-based approach of service providers, which are called to guarantee specific performance levels by fixing some constraints. When such constraints are violated, the SLA must be renegotiated, implying that the MUGF method has to be run again in order to return the best configuration guaranteeing the new performance level.

\section{Conclusion}

We propose an availability assessment approach to fit the modern Service Function Chain paradigm adopting: a Multi-State System model to represent the complex hardware and software stack in Containerized Network Functions; a queueing model, to include latency aspects; an extended version of multidimensional UGF technique, to efficiently analyze an infrastructure running several CNFs over multiple tenants, by combining their steady-state probability distributions through algebraic procedures. 

We used an experimental fault-injection testbed to estimate parameters for the model, such as the repair rates of CNF layers and the service rates of containerized instances. 
The proposed approach allowed us to efficiently solve the availability optimization problem within few minutes.

This work might be extended in several directions such as: \textit{i)} considering more and different service chains (e.g. mobile/broadband networks, data center chains) where the network manager is typically interested at finding the best redundant chain configuration at a minimal cost; \textit{ii)} differentiating (by priority or importance) the requests entering a chain, so as to deploy a service chain able to satisfy Quality-of-Service constraints, as well.

\balance

\vspace{-1cm}
\begin{IEEEbiographynophoto}{Luigi De Simone} (Ph.D.) is a postdoctoral researcher at the University of Naples Federico II, Italy. His research interests include dependability benchmarking, fault injection testing, virtualization reliability and its application on safety-critical systems.
\end{IEEEbiographynophoto}
\vspace{-1cm}
\begin{IEEEbiographynophoto}{Mario Di Mauro} (Ph.D.) is an assistant professor at the University of Salerno, Italy. His main fields of interest include: network performance, network security and availability characterization, data analysis for novel telecommunication infrastructures.
\end{IEEEbiographynophoto}
\vspace{-1cm}
\begin{IEEEbiographynophoto}{Roberto Natella} (Ph.D.) is an assistant professor at the University of Naples Federico II, Italy. His research interests include dependability benchmarking, software fault injection, software aging and rejuvenation, and their application in OS and virtualization technologies.
\end{IEEEbiographynophoto}
\vspace{-1cm}
\begin{IEEEbiographynophoto}{Fabio Postiglione} (Ph.D.) is an associate professor of statistics at the University of Salerno, Italy. His research interests include statistical characterization of degradation processes, reliability and availability modeling of complex systems, and Bayesian methods.
\end{IEEEbiographynophoto}

\onecolumn

\section*{Appendix A}	
We introduce two operators (namely series and parallel structure functions) to formally derive $\bm{\Delta}^{(m)}(t)$ and $\bm{\Delta}^{c}(t)$.
For the sake of simplicity, let us start to define the \textit{parallel structure function}:
\beq
\psi_p: \Omega^{L_m} \rightarrow \mathbb{R}^K \cup (+\infty,\dots,+\infty).
\eeq
Accordingly, the mean CSD introduced by tier $m\in \{P, S, I, H\}$ is:
\begin{align}
\nonumber
 & \bm{\Delta}^{(m)}(t) = \psi_p \left(   \boldsymbol{X}^{(m,1)}(t), X_{D}^{(m,1)}(t), X_{I}^{(m,1)}(t) , \dots,   \right. \\  
& \left. \boldsymbol{X}^{(m,L_m)}(t), X_{D}^{(m,L_m)}(t), X_{I}^{(m,L_m)}(t) \right)  = \left(\Delta_1^{(m)}(t),..., \Delta_K^{(m)}(t)\right), 
\label{eq:strfunpar}
\end{align}
where $\boldsymbol{X}^{(m,L_m)}(t)$ denotes the $\Omega_S$-valued failure/repair process of the Software layer, $X_{D}^{(m,L_m)}(t)$ denotes the $\Omega_D$-valued failure/repair process of the Docker layer, and $X_{I}^{(m,L_m)}(t)$ denotes the $\Omega_I$-valued failure/repair process of the Infrastructure layer.
Finally, $\Delta_i^{(m)}(t)$ is the stochastic process describing the $M/G/G_i^{(m)}(t)$ queue, that can be computed like in Section \ref{sect:queuemodel}, by replacing $\gamma \eta_i$ with $G_i^{(m)}(t) =  \sum_{\ell=1}^{L_m} G^{(m,\ell)}_{i}(t)$ in equations from (\ref{eq:sess}) to (\ref{eq:kingman}). 


It is now useful to recall that, since the call flow traverses the cIMS chain, the overall mean CSD is the sum of mean CSDs introduced by each single tier. 

Accordingly, by introducing $L_{\mathrm{tot}}= \sum_{m\in \{P, S, I, H\}} L_m$, we define the \textit{series structure function}:
\beq
\psi_s: \Omega^{L_{\mathrm{tot}}} \rightarrow \mathbb{R}^K \cup (+\infty,\dots,+\infty).
\eeq
Thus, the overall mean delay $\boldsymbol{\Delta^c}(t) = \left( \Delta_1^{c}(t),..., \Delta_K^{c}(t) \right)$ introduced by the cIMS is given by:
\begin{align}
\nonumber
 \bm{\Delta}^{c}(t) = \sum_{m\in \{P, S, I, H\}}\bm{\Delta}^{(m)}(t) = \psi_s \left(  \boldsymbol{X}^{(P,1)}(t), X_{D}^{(P,1)}(t), X_{I}^{(P,1)}(t) , \dots,   \right.   
\nonumber
 \left. \boldsymbol{X}^{(P,L_P)}(t), X_{D}^{(P,L_P)}(t), X_{I}^{(P,L_P)}(t) , \dots, 
  \right. \\ 
  \nonumber
 \left. \boldsymbol{X}^{(H,1)}(t), X_{D}^{(H,1)}(t), X_{I}^{(H,1)}(t) , \dots,
  \right. 
    \nonumber
   \left.  \boldsymbol{X}^{(H,L_H)}(t), X_{D}^{(H,L_H)}(t), X_{I}^{(H,L_H)}(t) \right)   \\ = 
    \sum_{m\in \{P, S, I, H\}} \psi_p  \left( \boldsymbol{X}^{(m,1)}(t), X_{D}^{(m,1)}(t), X_{I}^{(m,1)}(t) , \dots,   \right. 
   \left.  \boldsymbol{X}^{(m,L_m)}(t), X_{D}^{(m,L_m)}(t), X_{I}^{(m,L_m)}(t) \right).
\label{eq:strfunser}
\end{align}
%
%
%


\onecolumn
\section*{Appendix B}	


Let $p_1(t),\dots,p_{12}(t)$ be the state probabilities corresponding to states $S_1,\dots,S_{12}$, and $p_I(t)$ and $p_D(t)$ the state probabilities corresponding to states S$_I$ and S$_D$, respectively, as shown in Fig. \ref{fig:miniseph}.

According to ($6$), by assuming one and the same model for each CNF composing the cIMS system, all the state probabilities at time $t$ can be derived by solving the system of $14$ differential equations (\ref{eq:sistema}), representative of the $14$-state MSS in Fig. \ref{fig:miniseph}, with the constraint $\sum_{i=1}^{12} p_i(t) + p_D(t) + p_I(t) = 1$, and the assumption that the node is initially working. 
\begin{figure*}[h]
					\begin{equation}
						\begin{aligned}
							\newcommand{\lamuno}{\lambda_{C_1}}
							\newcommand{\lamdue}{\lambda_{C_2}}
							\newcommand{\lamd}{\lambda_D}
							\newcommand{\lami}{\lambda_I}
							\newcommand{\muuno}{\mu_{C_1}}
							\newcommand{\mudue}{\mu_{C_2}}
							\newcommand{\mud}{\mu_D}
							\newcommand{\mui}{\mu_I}
							\left\{
							\begin{array}{l c l}
								\displaystyle{\frac{dp_I(t)}{dt}} &=& -\mui p_I(t) + \lami \sum\nolimits_{i=1}^{12}p_i(t) + p_D(t) \\[10pt]
								\displaystyle{\frac{dp_D(t)}{dt}} &=& -(\mud+\lami) p_D(t) + \lamd \sum\nolimits_{i=1}^{12}p_i(t) \\[10pt]
								\displaystyle{\frac{dp_1(t)}{dt}} &=& -(2\muuno + 3\mudue + \lamd +\lami)p_1(t) +\lamuno p_2(t) + \lamdue p_3(t) \\[10pt]
								\displaystyle{\frac{dp_2(t)}{dt}} &=& 2\muuno p_1(t)  - (\lamuno+ \muuno + 3\mudue + \lamd + \lami)p_2(t) + 2\lamuno p_4(t) + \lamdue p_5(t) \\[10pt]
								\displaystyle{\frac{dp_3(t)}{dt}} &=& 3\mudue p_1(t) - (\lamdue+ 2\muuno+2\mudue + \lamd + \lami)p_3(t) +\lamuno p_5(t) + 2\lamdue p_6(t) \\[10pt]
								\displaystyle{\frac{dp_4(t)}{dt}} &=& \muuno p_2(t) - (2\lamuno + 3\mudue + \lamd + \lami)p_4(t) + \lamdue p_7(t) \\[10pt]
								\displaystyle{\frac{dp_5(t)}{dt}} &=& 3\mudue p_2(t) + 2\muuno p_3(t) - (\lamuno+\lamdue+ \muuno + 2\mudue +\lamd + \lami)p_5(t) + 2\lamuno p_7(t) + 2\lamdue p_8(t) \\[10pt]
								\displaystyle{\frac{dp_6(t)}{dt}} &=& 2\mudue p_3(t) - (2\lamdue+2\muuno+\mudue+\lamd + \lami)p_6(t) + \lamuno p_8(t) + 3\lamdue p_9(t) \\[10pt]
								\displaystyle{\frac{dp_7(t)}{dt}} &=& 3\mudue p_4(t) + \muuno p_5(t) -(2\lamuno+\lamdue+2\mudue+\lamd + \lami)p_7(t) + 2\lamdue p_{10}(t) \\[10pt]
								\displaystyle{\frac{dp_8(t)}{dt}} &=& 2\mudue p_5(t) + 2\muuno p_6(t) -(\lamuno+2\lamdue+\muuno + \mudue+\lamd + \lami)p_8(t) + 2\lamuno p_{10}(t) + 3\lamdue p_{11}(t) \\[10pt]
								\displaystyle{\frac{dp_9(t)}{dt}} &=& \mudue p_6(t) - (3\lamdue + 2\muuno +\lamd + \lami)p_9(t) +\lamuno p_{11}(t) \\[10pt]
								\displaystyle{\frac{dp_{10}(t)}{dt}} &=& 2\mudue p_7(t) + \muuno p_8(t) -(2\lamuno+2\lamdue+\mudue+\lamd + \lami)p_{10}(t) + 3\lamdue p_{12}(t) \\[10pt]
								\displaystyle{\frac{dp_{11}(t)}{dt}} &=& \mudue p_8(t) + 2\muuno p_9(t) -(\lamuno+3\lamdue+\muuno+\lamd + \lami)p_{11}(t) + 2\lamuno p_{12}(t) \\[10pt]
								\displaystyle{\frac{dp_{12}(t)}{dt}} &=& \mudue p_{10}(t) + \muuno p_{11}(t) + \mud p_D(t) + \mui p_I(t) -(2\lamuno+3\lamdue+\lamd + \lami) p_{12}(t) %
							\end{array}
							\right.
						\end{aligned}
						\label{eq:sistema}
					\end{equation}
					\label{eq:supereqdiff}
\end{figure*}

The steady-state probability distribution $\bm{p}$ of the CNF in Fig. \ref{fig:miniseph} can be hence derived by considering the limit for $t\rightarrow \infty$ of the solution of system (\ref{eq:sistema}).
Alternatively, $\bm{p}$ can be can be simply determined by nullifying the derivative terms in system (\ref{eq:sistema}), and solving it along with the constraint $\sum_{i=1}^{12} p_i + p_D + p_I = 1$.


\end{document}